\documentclass[12pt]{iopart}
\usepackage{iopams,amstext,enumerate,graphicx}  

\newcommand{\Prob}{\mathbb{P}}

\newcommand{\ignore}[1]{}

 \newcommand{\fig}[1]{Fig.~\ref{#1}}
\newcommand{\sect}[1]{Sec.~\ref{#1}}
\newcommand{\Eq}[1]{Eq.\ (\ref{#1})} \newcommand{\Fig}[1]{Fig.~\ref{#1}}
 
\newcommand{\hmin}{ h_{\textrm{min} } }

\begin{document}

\title{RNA Secondary Structures: Complex Statics and 
Glassy Dynamics}

\author{S  Wolfsheimer} 
\address{ Institut f\"ur Physik, Universit\"at Oldenburg, Germany}
\ead{wolfsh@theorie.physik.uni-oldenburg.de}

\author{B Burghardt} 

\address{
  Institut f\"ur Theoretische Physik, Universit\"at G\"ottingen,
  Germany}
\ead{burghard@physik.uni-goe.de} 

\author{A Mann} 
\address{ Institut f\"ur Theoretische Physik, Universit\"at
  G\"ottingen, Germany}
\ead{mann@theorie.physik.uni-goettingen.de}

\author{A K Hartmann} 
\ead{a.hartmann@uni-oldenburg.de}
\address{ Institut f\"ur Physik, Universit\"at Oldenburg, Germany}

%\date{\today}

\begin{abstract} Models for RNA secondary structures (the topology of
folded RNA) without pseudo knots are disordered systems with a complex
state-space below a critical temperature.  Hence, a complex  dynamical
(glassy) behavior can be expected, when performing Monte Carlo 
simulation. Interestingly, in contrast to most other complex systems, 
the ground states and
the density of states can be computed in polynomial time exactly using
transfer matrix methods.  Hence, RNA secondary structure is an ideal
model to study the relation between static/thermodynamic properties
and dynamics of algorithms. Also they constitute  an ideal benchmark
system for new Monte Carlo methods.

  Here we considered three different recent Monte Carlo approaches: entropic
sampling using flat histograms, optimized-weights ensembles, and ParQ, which
estimates the density of states from transition matrices.
    These methods were examined by comparing the obtained density of 
states with
the exact results.  We relate the complexity seen in the dynamics of the
Monte Carlo algorithms
to static properties of the phase space by studying
the correlations between tunneling times, sampling errors, amount 
of meta-stable states and degree of ultrametricity at finite temperature.
\end{abstract}

\maketitle

\section{Introduction} 

One fundamental question concerning disordered systems is whether and how
static properties like the occurrence of phase transitions, 
low-energy meta-stable states and ultrametricity of the energy landscape
are related to dynamical properties like glassiness, existence of aging 
phenomena and computational hardness 
\cite{Young1998,Binder2005,Hartmann2005,Henkel2007}. 
Since almost all interesting systems cannot be treated analytically, one
has to resort to numerical approaches. Unfortunately,
those systems which
exhibit glassy behavior are usually also difficult to study numerically in
equilibrium, because the glassiness itself prevents reaching the equilibrium
during Monte Carlo (MC) simulations
for even moderate systems sizes. Hence, usually it is very 
difficult to study the above mentioned relation.

Here, we study a model \cite{Higgs1996} of RNA secondary structures,
which describe the biological and physical properties of RNA already very well
\cite{Tinoco1999}.
The simple model investigated here
already has been proven to be very useful to understand low-temperature
properties of RNA see e.g.
\ Refs.\cite{Higgs1996,Pagnani2000,Hartmann2001,Bundschuh2002,Krzakala2002,Burghardt2005} and most recently
\cite{Lassig2006}. 
 
Interestingly,
besides its usefulness for molecular biophysics, this model is of
fundamental interest to understand the  relation between static and 
dynamic properties of disordered systems:
\begin{itemize}
\item The model exhibits quenched disorder and
 has a complex low-energy landscape, where an interesting dynamical
behavior can be expected.

\item  The model exhibits a static phase transition at finite temperature
 between a ``homogeneous'' high-temperature phase and a ``complex'' 
low-temperature phase. The latter phase exhibits
an ultrametric  phase-space structure \cite{Higgs1996}. 

\item The static behavior of the model can be analyzed exactly using
 partition-function calculations for each single
realization of the disorder. The
computation time grows only  polynomially with system size.
This approach also allows to generate secondary-structure
configurations in equilibrium without
rejection and exhibiting zero correlations between different configurations.
\end{itemize}
Only a few models which combines all three properties in one are known. 
For example
two-dimensional $\pm J$ Ising spin glasses and fully frustrated models 
can be solved exactly by transfer matrix methods \cite{Morgenstern1980} or
by the program of Saul and Kardar in polynomial time \cite{Saul1994}.
On the other hand, no rejection-free equilibrium sampling method
is known. Furthermore,  two-dimensional spin glasses
only have a  phase transition at zero temperature \cite{Villian1977}.
Better comparable to the RNA secaondary structures is
a model of directed polymers in random media \cite{Mezard1990}, where
direct sampling using transfer matrices of the partition function could
be used and a non-trivial phase transition was detected.

On the algorithmic side, in the nineties of the last century a lot of
technical advances in computer technology had been made.  This allows
for computational investigations of larger physical systems with
smaller errorbars. Hence, finite-size-scaling methods, which usually
need simulation over some orders of magnitude in system size, become
more accurate.

In parallel to the technical advances a number of new Monte Carlo (MC)
algorithms, which go beyond standard Metropolis sampling
\cite{Metropolis1953}, had been developed.  Very popular are
algorithms for calculating the density of states (DOS) $g(E)$ of a
physical system, because with the knowledge of this quantity the
canonical partition function $Z(T)$ is known for all temperatures and
the problem is solved in principle.

A wide class of such algorithms are so called generalized ensemble
algorithms. Entropic sampling \cite{Lee1993}, multicanonical sampling
\cite{Berg1992} and most recently Wang-Landau sampling
\cite{Wang2001,Wang2001a} are only a few examples.  All these
algorithms are based on the idea of importance sampling,
where the probability of interest is modified in such a way, that
``rare events'' become more probable and the system may overcome
typical barriers.  Trebst et al.\ \cite{Trebst2004} introduced an
iterative algorithm to optimize the round-trip time. Similar
algorithms are also available for simulated and parallel tempering
\cite{Geyer1995,Katzgraber2006,Rathore2005}.
Another method, ParQ \cite{Heilmann2005}, is based on approximating
the infinite temperature transition matrix from a non-equilibrium
simulation.

In this article we will study the relationship of static and dynamic 
properties of RNA secondary structures. 
We relate the complexity seen in the dynamics of the three mentioned
Monte Carlo approaches
to static properties of the phase space by studying
the correlations between tunneling times, sampling errors, amount
of meta-stable states and ultrametricity.
Note that this kind of dynamics does not allow for sampling real biological 
processes such as kinetic folding paths (see for example \cite{Isambert2000}).

The article is organized as follows. After a brief introduction to the
model and exact transfer matrix approach in \sect{sec:RNA} we review
the general idea of estimating the infinite temperature transition
matrix, which is independent of the particular method to access the
configuration space.  Then we will specify the three methods, we used
and discuss their advantages and disadvantages.  In
\sect{sec:convergence} we will consider
convergence properties of the MC approaches and
relate the algorithmic complexity of the MC approaches to
structural complexity of the phase space.  In particular we will focus
on a possible reason for the failure of the ParQ algorithm in our case.

%%%%%%%%%%%%%%%%%%%%%%%%%%%%%%%%%%%%%%%%%%%%%%%%%
%                                               %
% RNA Secondary Structure                       %
%                                               %
%%%%%%%%%%%%%%%%%%%%%%%%%%%%%%%%%%%%%%%%%%%%%%%%%
\section{RNA Secondary Structure}
\label{sec:RNA}
RNAs are a biological macromolecules consisting of linear chains built
from four types of bases: adenine (A), cytosine (C), guanine (G) and
uracil (U). Each molecule is characterized by its fixed sequence of bases,
the so called {\em primary structure}, in the same way as for DNA.
 In contrast to DNA, RNAs are single stranded molecules and
therefore may build hydrogen bonds within the same chain (RNA
``folding'').  The relevant
description of RNAs is the {\em secondary structure}  i.e.\ 
 the set of pairs of bases connected each by hydrogen bonds.   Finally, the
{\em tertiary structure} describes to the spacial structure of the
folded molecule and is considered less important for an RNA's behavior
than the secondary structure \cite{Tinoco1999}. Note that this is
 in contrast to proteins, where the tertiary structure is most important.

Let us denote the alphabet of all possible bases as $\Sigma =
\left\{A,C,G,U\right\}$ and a RNA sequence of length $L$ by
$\mathcal{R}=r_1\,r_2\,\cdots\,r_L$, where $r_i \in \Sigma$.  A
secondary structure of $\mathcal{R}$ is a set   $\mathcal{S}$ of 
pairs $(i,j)$, where $i,j\in\left\{1,\dots,L\right\}$ and with the convention
$i<j$. $N=|\mathcal{S}|$ denotes the
number of pairs. In the following, we frequently just write ``structure'',
when we refer to secondary structures.

Because any base can be paired only with one other base, each
base number can appear at in at most one pair of $\mathcal{S}$. There
are in
principle three possible cases for two pairs $(i,j)$ and $(k,l)$,
namely
\begin{enumerate}[(a)]
\item{ $i < j < k < l$ }
\item{ $i < k < l < j$ }
\item{ $i < k < j < l$ (pseudo knots) }.
\end{enumerate}

\begin{figure}
  \begin{center}
    \includegraphics[clip,width=0.95\textwidth]{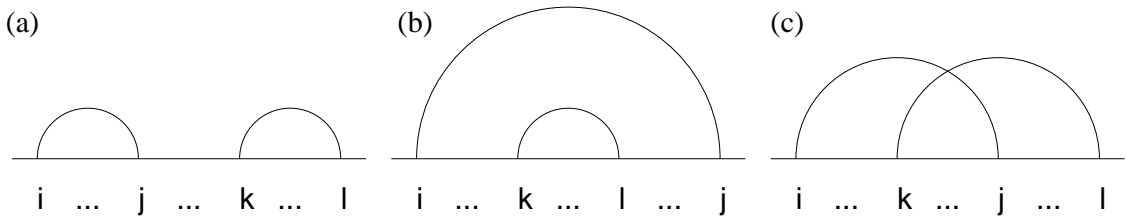}
    \caption{
      \label{fig:STRUCT} 
      Three different cases of orders of the pairs $(i,j)$ and $(k,l)$
      illustrated in the arc representation of RNA secondary
      structure: (a) $i < j < k < l$, where the bonds are completely
      separated.  (b) $i < k < l < j$, where $(i,j)$ includes $(k,l)$.
      (c) $i < k < j < l$, the case of pseudo knots.  }
  \end{center}
\end{figure}

These three cases are illustrated in \fig{fig:STRUCT}, where the
sequence is represented by a horizontal axis and arcs between
positions describe pairs.  
Here we follow the notion of pseudo knots being more an element of the 
tertiary structure \cite{Tinoco1999} 
and neglect them, i.e.\ only cases (a) and (b) of
\fig{fig:STRUCT}  can appear in the structures.

Due to the bending rigidity of the RNA molecule it is impossible that two
bases $r_i$ and $r_j$ close to each other in the primary structure can be
paired, therefore we require a minimum distance, i.e.\ $j-i \geq \hmin$, here
we use $\hmin = 2$.

Chemically, 
the base pairs adenine - uracil are formed by two and cytosine - guanine are
formed by three
hydrogen bonds.  Pairs of bases that may form bonds are said to be
complementary (A-U and C-G).  We used a simple energy model, which
assigns an energy to each possible pair of bases $a,b\in\Sigma$
\begin{equation}
e(a,b) =
\cases{
  -e_p & if $a$, $b$ are complementary  \\ \infty &  otherwise ,
}
\end{equation}
$e_p > 0$ is a tunable parameter and set to $1$ in this study.  The
energy of the structure $\mathcal{S}$ for sequence ${\mathcal{R}}$ 
is defined by
\begin{equation}
E(\mathcal{R},\mathcal{S}) = \sum_{(i,j)\in\mathcal{S}} \, e(r_{i},r_{j}).
\label{eq:energy}
\end{equation}

%%%%%%%%%%%%%%%%%%%%%%%%%%%%%%%%%%%%%%%%%%%%%%%%%%%%%%%%%%%%%%%%%%%%%%
%                                                                    %
% Exact calculations of the partition function and density of states %
%                                                                    %
%%%%%%%%%%%%%%%%%%%%%%%%%%%%%%%%%%%%%%%%%%%%%%%%%%%%%%%%%%%%%%%%%%%%%%

\subsection{Exact calculations of the partition function and density of states}
The study of
RNA secondary structures without pseudo knots falls into the class of
problems, that can be treated exactly by transfer matrix methods (or
dynamic programming) because it can be formulated recursively 
\cite{Nussinov1978,Zuker1989,Gennes1968}.

The first quantity of interest here is the canonical partition
function.  There are $L^2/2 - L/2 $ possible subsequences $r_i\,\cdots
r_j$ ($i < j$) with corresponding partition functions $Z_{i,j}$.  The partition
function of $Z_{i,j+1}$ depends on all $Z_{k,l}$ ($i\le k<l\le j$)
and the subsequence $r_i\,\cdots r_j\,r_{j+1}$ only. One has to sum
over the different cases for the last position $j+1$. 
There are at most $j-i-\hmin$ candidate pairs that connect the
base at the position ${j+1}$ with any other base at
position $k$ in the subsequence. Due to the definition of the energy
model positions with $j-k \le \hmin$ and non-complementary bases
are excluded. Since pseudo knots are also excluded, the hypothetically
inserted pair $(k,j+1)$ induces two independent subsystems $r_i\cdots
r_{k-1}$ and $r_{k+1} \cdots r_{j}$.  Hence the partition function
$Z_{i,j+1}$ is given by
\begin{equation}
\label{eq:dynprogr}
Z_{i,j+1} = Z_{i,j} + \sum_{k=i}^{j-\hmin}\, Z_{i,k-1}\,\mathrm{e}^{-\beta
e(r_k,r_{j+1})}\, Z_{k+1,j}
\end{equation}

With $\beta = 1/T$, $k_B=1$ and according to our energy model the
factors $\mathrm{e}^{-\beta e(r_k,r_{j+1})}$ are given by
\[
\mathrm{e}^{-\beta e(r_k,r_{j+1})} = 
\cases{
  \mathrm{e}^{\beta}  &  if  $r_k$ and $r_{j+1}$ are complementary \\ 
  0 & otherwise\\}
\]

Starting with the boundary conditions $Z_{i,i} = 1$ and $Z_{i,i-1} = 1$, one
can calculate $Z_{i,j}$ for increasing values of $j-i$, finally arriving
at $Z_{1,L}$ which yields the full partition function.  Since the
number of possible sequences grows quadratic in the sequence length
and Eq. \ref{eq:dynprogr} can be computed in linear time, the overall
time complexity is of order $L^3$ and the required memory grows
like $L^2$.

\begin{figure}
  \begin{center}
    \includegraphics[clip,width=0.97\linewidth]{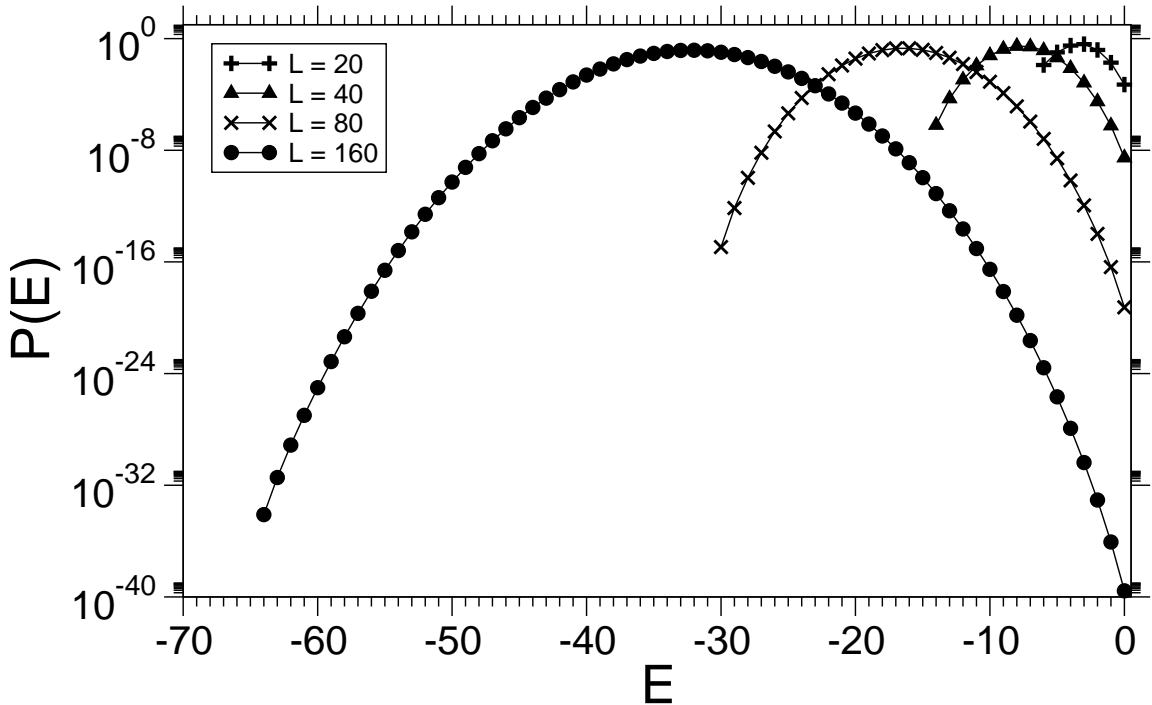}
    \caption{
      \label{fig:DOS} 
      Normalized DOS of different randomly generated RNA
    sequences. Lines are guides to the eyes only.  }
  \end{center}
\end{figure}

The dynamic programming algorithm for the partition function can be
easily generalized to exact DOS calculations  in
$\mathcal O(L^5)$ time complexity.
This is possible because the energy occurs in multiples of an energy
``quantum'' (the energy of a pair). Then the canonical partition
function at inverse temperature 
$\beta=\ln z$ of the subproblems $r_i
\cdots r_j$ can be written as a polynomial in $z$: $\tilde{Z}_{i,j}(z) =
\sum_{n=0}^{n_{\max}} g_{i,j}(E_n) \cdot z^n$.  In this notation the
index $n$ is the number of pairs and hence, using our simple energy
model, $E_n = -e_p n = -n$.  The coefficients $ g_{i,j}(E_n)$ 
of this polynomials yield the DOS
of the subproblems and the full DOS is given by $g(E_n) =
g_{1,L}(E_n)$.

The generalization of \Eq{eq:dynprogr} is given by

\begin{eqnarray*} 
  &\tilde{Z}_{i,j+1}(z) &=  \tilde{Z}_{i,j}(z) +  
  \sum_{k=i}^{j-\hmin}\, \tilde{Z}_{i,k-1}(z)\,
  %\mathrm{e}^{-\beta   e(r_k,r_{j+1})}
  c_{r_k,r_{j+1}}(z)
  \, \tilde{Z}_{k+1,j}(z)\\
\text{where}\quad&
  c_{r_k,r_{j+1}}(z)&=\cases{
  z  &  if  $r_k$ and $r_{j+1}$ are complementary \\ 
  0 & otherwise}
\end{eqnarray*}

with similar boundary conditions $\tilde{Z}_{i,i}(z) = 1$ and
$\tilde{Z}_{i,i-1}(z) = 1$.  Since the numbers $Z_{i,j}$ are replaced
by polynomials and multiplications between these objects are involved,
the time complexity increases to $\mathcal{O}(L^5)$. In \Fig{fig:DOS}
the normalized DOS $P(E)=\frac{1}{\sum_{E'} g(E')} \cdot g(E) $ for
four different randomly generated RNA sequences of lengths between
$20$ and $160$ are illustrated. Note that the DOS depends on the
disorder realization, i.e.\ on the sequence and that the DOS can be
obtained exactly over the full range of energy values, easily 
resulting in normalized densities of $10^{-40}$. Hence, RNA secondary
structure is a complex-behaving model with quenched disorder that
can be treated exactly for each realization. This makes it very valuable
for the study of the dynamics of algorithms, like MC approaches,
as considered in the next section.

%%%%%%%%%%%%%%%%%%%%%%%%%%%%%%%%%%%%%%%%%%%%%%%%%%%%%%%%%%%%%%%%%%%%%%%%%%%%%%%%%%%%%%%%%%%%%%%
%
% Monte Carlo algorithms 
%
%%%%%%%%%%%%%%%%%%%%%%%%%%%%%%%%%%%%%%%%%%%%%%%%%%%%%%%%%%%%%%%%%%%%%%%%%%%%%%%%%%%%%%%%%%%%%%%
\section{Monte Carlo methods to obtain the DOS}

In this section we review the MC algorithms used in this
study.  All algorithms are based on a Markov chain in the space of
structures with a fixed sequence ${\mathcal R}$,
starting with an empty structure. For simplicity, the dependence
of all quantities on ${\mathcal R}$ is not stated explicitly below.
Before explaining the algorithms in details, a few general statements are
made, which apply to all algorithms studied here.

\subsection{Three classes of proposals}

Given a structure $\mathcal{S}$, a single MC proposal
consists of selecting a candidate structure $\mathcal{T}$ from its
local neighborhood.  The neighborhood of $\mathcal{S}$ is defined here by
a removal or insertion of a bond, i.e.\ two kind of MC moves
are imposed.  The proposal is accepted according to the Metropolis
acceptance rule
\begin{equation}
  \label{eq:metropolis}
  r = \min\left\{1, w(E(\mathcal{T} )) / w( E( \mathcal{S} ))
  \right\},
\end{equation}
where $w(\cdot)$ is a weight function which depends on the
energy $E(\cdot)$ of a structure given by Eq.\ (\ref{eq:energy}).
This ratio of weights between the new and the old structure is
implementation dependent.  For example it depends on the temperature
in the canonical ensemble or on the weights of a generalized ensemble.

At the beginning of the simulation a list of all $N_{\text{possible}}$
\emph{possible pairs} $\{(i,j)\}$ is created.
These pairs are compatible to the energy model ($e(r_i,r_j)<\infty$)
and have sufficient distance $\hmin$ along the sequence.
Note that there are $\mathcal{O}(L^2)$ of possible pairs.

This class of pairs can be divided into three classes at each stage of
the simulation in terms of the \emph{n-fold way} \cite{Bortz1975}.  The first
class are the \emph{active pairs}, i.e.\ that pairs that are
currently bonded.  The class of inactive pairs can be divided into two
subclasses, namely \emph{allowed pairs}, which can be inserted to the
current structure according to the energy model, and \emph{currently 
forbidden pairs}, which would lead to
pseudo knots for example. Structures which include forbidden pairs 
(which will never occur in our simulations), we call forbidden as well. 
Formally, structures, which have forbidden pairs, have zero weight in
\Eq{eq:metropolis}.

Active pairs are associated with an energy change of $\Delta E = 1$,
allowed pairs with $\Delta E = -1$ and forbidden pairs with $\Delta E
= \infty$.  The number of members in the classes given structure
$\mathcal{A}$ are denoted by $N( \mathcal{A},+1)$, $N(\mathcal{A},-1)$
and $N(\mathcal{A},0)$ for active, possible and forbidden pairs
respectively.

A secondary structure is represented as list of links to the static
array of possible pairs. Then the simulation requires some bookkeeping
of the lists for all three classes. For this purpose it makes sense to
setup a list of cross-links between all pairs indicating
incompatibility, i.e.\ a list of pairs of pairs that lead to
pseudo knots, when they both are inserted at the same time.

The most simple approach, ensuring detailed balance, would work in
the following way. For each MC step, select one pair among
all possible pairs. Then
\begin{itemize}
\item If the pair is allowed or active, insert or remove (i.e.\ ``flip'') it 
from the structure with probability given by (\ref{eq:metropolis}).
Hence, for standard MC, an allowed pair would be inserted always, and
an active pair removed with some probability smaller than one.
\item If the pair is forbidden, do nothing.
\end{itemize}

However in ``dense'' systems this algorithm gets stuck very quickly
because in many cases a forbidden pair would be selected in almost all
cases, because in this case the number of active and allowed pairs is
$\mathcal O(L)$
only.
  In order to speed up this algorithm, we use a dynamics, that avoids
attempts leading to forbidden pairs and selects pairs from the list of
active or possible pairs only. Each of the $N(\mathcal{A},+1) +
N(\mathcal{A},-1)$ pairs have the same probability to be selected.
The forbidden attempts are taken into account, by advancing the 
simulation-time clock sufficiently. 
This kind of dynamics combines a ``rejection-free dynamics'', as
implemented in the {n-fold way} \cite{Bortz1975}, with standard
acceptance probabilities. Details are presented now.

%We will denote this kind of dynamics as ``semi rejection-free''.
The dynamics of an algorithm can be seen as a random walk
in structure (configuration/state) space.
When performing the simulation, one has to account for the \emph{waiting
times} $m$, that the random walker would have stayed in the structure
$\mathcal{S}$ due to forbidden transitions in the local environment.  It
can be computed by following considerations: Let $p$ be the
probability that a forbidden pair is selected, given that the random
walker sits in state $\mathcal{A}$, i.e.\  $ p = N(\mathcal{A},0) /
N_{\text{possible}}.  $ Then probability that the random walk selects
a non-forbidden pair in the
the current state after $m$ trials is given by
\begin{equation}
\label{eq:geometrical-distribution}
p(m) = p^m (1-p).
\end{equation}
The probability of staying at most $\tau$ time steps can be evaluated
via geometric progression:

\begin{equation}
\label{eq:waiting-distribution}
P(\tau) = \Prob\left[ m \leq \tau \right] = \sum_{m=0}^\tau p^m \cdot
(1-p) = 1 - p^{\tau+1}
\end{equation}

In order to assign a waiting time to the current structure one has to draw a
random number according to the discrete distribution
\Eq{eq:waiting-distribution}, i.e.
\[
\tau = \lfloor \ln(\zeta) / \ln(p) \rfloor,
\]  
where $\zeta$ is an uniformly distributed random number and $\lfloor x
\rfloor$ denotes rounding down to the next integer.

After the simulation-time clock has advanced by the waiting time, 
a pair is selected from the set of
active and allowed pairs with uniform probability, and the
pair is flipped with a probability given by (\ref{eq:metropolis}).
Hence, if the flip is rejected, then the current structure persists.
This completes one MC step.

There are two timescales involved. The \emph{computer time} measures
the number of MC steps and the \emph{MC time} is
associated with a physical time scale of the random walker.  The first
plays the major role in the performance analysis of the algorithms and
the latter one gives the correct weight to the visited states.

Next we want to sample three quantities of interest, the
\emph{energy}, a random \emph{energy change} $\Delta E$ associated with
each attempt (regardless if the step is accepted or not) and  a random
\emph{waiting time}.  To sum it up, we want to sample the sequence
\[
(\hat{E}_1, \Delta\hat{E}_1, \hat{\tau}_1),\ldots,(\hat{E}_{m},
\Delta \hat{E}_m, \hat{\tau}_m).
\]

In the following we will outline the two MC schemes, that were
used to collect data for our estimate. 
In both cases, we study the evolution on the
scale of \emph{macrostates}, which contain all structures of the same
energy.
The first approach is based
one generating flat histograms for the macrostates, while the second
approach estimates the DOS from transition matrices.

\subsection{Flat-histogram ensemble}
Instead of sampling configurations according to the Boltzmann weight
$w(E) \propto \exp(-\beta E)$, flat histogram methods use a different
weight function, such that all macrostates are visited with equal
probability. A \emph{perfect flat histogram} ensemble, where $w(E)
\propto 1/g(E)$, requires the knowledge of the DOS
$g(E)$. As already outlined, many advances had been made to estimate the DOS
 iteratively, for example Wang-Landau sampling
\cite{Wang2001,Wang2001a}.  In a final stage of this algorithm one
usually fixes a good approximation of the DOS and performs a production run
in an almost flat histogram ensemble.  Here we take advantage of the fact
that we can obtain the exact DOS for each realization of the
disorder. This allows us to 
measure the performance of the perfectly flat histogram
ensemble, which results in  an upper bound for other approaches, 
where an exact DOS calculation is not possible.

\subsection{Optimized flat-histogram ensemble}
Perfectly flat histogram methods are only optimal in the sense, that
all macrostates are visited with equal probability. There might
remain large correlations due to the fact that the random walker stays
in local minima for a long time. Especially near phase transitions,
where the specific heat diverges, a huge amount of computation time is
spent.  This effect is known as \emph{critical slowing down}.

This is also related to \emph{regeneration} of Markov chains in the
following sense: An Markov chain is regenerative if there are times
$T_i$, such that the process after $T_i$ becomes independent from
times before $T_i$. 
Obviously our model has a regeneration point, whenever the
random walk hits the null structure.  The path between regeneration
points are called \emph{tours}.  Usually the distribution of tour
lengths exhibits a heavy tail and only a very small fraction of tours hit
one of the ground states.  The \emph{first-passage time} (also called
\emph{tunneling time}) is the time the random walker needs to hit the
ground state starting at its last regeneration point. 
This is an extremal event and, hence the distribution of
first-passage times might be, at least approximately, a 
generalized extreme-value distribution exhibiting a heavy tail.

Small first-passage times
increase mixing and performance of the sampler.  We will also use 
\emph{round-trip time}, which is the tunneling time plus the time
needed to go back to regeneration.  Since the turn from regeneration
to the ground state is much longer than the turn back, first-passage time
and round trip equal approximately.

Trebst et al.\ \cite{Trebst2004} developed an iterative algorithm to
optimize round-trip times in a generalized ensemble.  Instead of giving
all macrostates the weight $w(E) \propto 1/g(E)$ a different weight
function $w^{\text{opt}}(E)$ is chosen, such that the \emph{ number
of round trips } between an energy interval $E_+$ and $E_-$ is
maximized.

The equilibrium distribution of the optimized ensemble is proportional
to $w^{\text{opt}}(E) \cdot g(E)$, which is not a flat histogram in
general.  The methods works for both, for Metropolis and n-fold
dynamics.  Consider the  steady state current from $E_+$ to $E_-$,
which is to first order
\[
j \propto D(E) \cdot w^{\text{opt}}(E) \cdot g(E) \frac{df}{dE}
\]
for a continuous energy. $D(E)$ denotes the local diffusivity at energy 
level $E$ and $f(E)$ is the fraction of visits at energy $E$, where the last visits 
to $E_+$ has occurred more recently than to $E_-$.
A sample estimate for $f$ can
be made by labeling the random walker with two different labels ``$+$''
or ``$-$'', depending on whether it has visited the state $E_+$ or $E_-$
most recently.  During the simulation a separate energy histogram
$H_\pm (E)$ for each label is updated and an approximation of $f$ for
the given weights is given by
\[
\hat{f}(E) = \frac{H_+(E)}{H_+(E) + H_-(E)}.
\]
The feedback iteration that converges to the optimized ensemble for
Metropolis updates is given by \cite{Trebst2004}
\begin{equation}
\label{eq:opt-iteration-metropolis}
w^{k+1}(E) = w^{k}(E) \cdot \sqrt{ \frac{1}{ H_+(E) + H_-(E)} \cdot
\frac{df}{dE} }.
\end{equation}
For the n-fold way, or as in our case the semi rejection-free
dynamics, the iteration has to account for the two intrinsic time
scales.  Since one aims at optimizing the computer time the iteration
scheme \Eq{eq:opt-iteration-metropolis} is modified by the factor
$\tau(E)$, which is the accumulated waiting time at energy level $E$,
in other words
\begin{equation}
\label{eq:opt-iteration-nfold}
w^{k+1}(E) = w^{k}(E) \cdot \sqrt{ \frac{1}{ H_+(E) + H_-(E)} \cdot
\frac{df}{dE} \cdot \frac{1}{\tau(E)} }.
\end{equation}
After each iteration the number of MC steps, which is used to accumulate
the histograms, is doubled. The
derivative $df/dE$ can be approximated by a polynomial interpolation
of $f(E)$ and numerical derivation.

\begin{figure}
  \begin{center}
    \includegraphics[clip,width=0.95\linewidth,angle=-0]{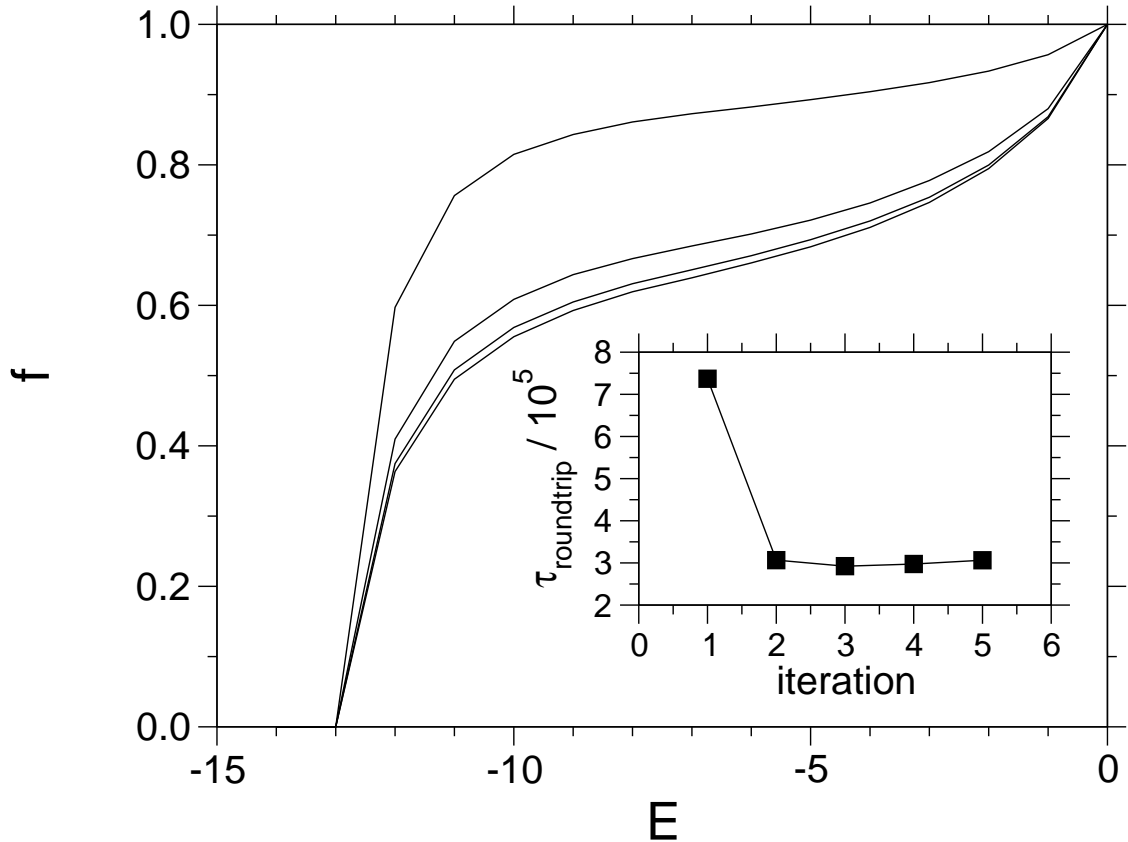}
    \caption{
      \label{fig:CONVOPT} 
      Convergence of $f(E)$ using iteration scheme
      \Eq{eq:opt-iteration-nfold}, shown by lines
      connecting the data points, for better visibility.  
      The energy weights were optimized
      between $E_- = E_1$ (first excitations) and the null structure
      $E_+ = 0$.  Between iteration 4th and 5th no significant
      difference of $f(E)$ is visible.  Inset: Convergence of 
      round-trip times. Lines in the inset are guides to the eyes only. }
  \end{center}
\end{figure}

Since the events for going from a first excited state $E_1$ to the
ground state $E_0$ occur very rarely, the statistics for $f$ might be
bad and the iteration scheme \Eq{eq:opt-iteration-nfold} converges
slowly, if the complete energy spectrum is considered for the
optimized ensemble. However, an optimization over the complete range is
not essential and it is sufficient to optimize the ensemble from first
excitations $E_- = E_1$ to the null structure $E_+ = 0$. The link to
the remaining weight $w^{i+1}(E_0)$ can be made by requiring the next
iteration to visit either the ground state or the first excitations with equal
probability (any other finite fraction will work as well), i.e.
\begin{eqnarray*}
  w^{i+1}(E_0)
  & = &w^{i+1}(E_1) \cdot \frac{g(E_1)}{g(E_0)} \\ &
  \approx & w^{i+1}(E_1) \cdot \frac{H_+(E_1) + H_-(E_1)}{H_+(E_0) + H_-(E_0)} \cdot
  \frac{w(E_1)}{w(E_0)}.
\end{eqnarray*}

Note that the random walker itself is not restricted to
$\left[E_-,E_+\right]$. During each iteration one can measure the
round-trip time of the random walker over the full spectrum from $E_0$
to the null structure as a quantity which describes the
performance. For a small system $L=40$ we compared the performance of
the optimization over the full spectrum and the restricted spectrum
and found no significant difference in round-trip times.  In both
cases the round-trip time decreases by a factor of about $2$ already
in the second iteration of updating the weights.  
The effect of decreasing round-trip times
becomes more pronounced for larger systems (see
\sect{sec:convergence}).

%%%%%%%%%%%%%%%%%%%%%%%%%%%%%%%%%%%%%%%%%%%%%%%%%%%%%%%%%%%%%%%%%%%%%%%%%%%%%%%
%
% TRANSITION MATRIX MONTE CARLO
%
%%%%%%%%%%%%%%%%%%%%%%%%%%%%%%%%%%%%%%%%%%%%%%%%%%%%%%%%%%%%%%%%%%%%%%%%%%%%%%%
\subsection{DOS through transition matrix estimates}

The method, we will examine next, is based on evaluating
 data from infinite temperature transition matrices.
Hence, it  has similarities to transition
matrix MC (TMMC), which was introduced by Lee and Wang in
several publications \cite{Wang1999,Wang2000,Wang1999a}.
In TMMC, one uses an acceptance rate which is more general than
\Eq{eq:metropolis}.  Here we keep the original weights for generalized
(flat histogram or optimized) ensemble or simulated annealing (ParQ)
dynamics with Metropolis update rules.

The connection between the transition matrix and DOS can be made as
follows:

Consider an infinitely long simulation in the canonical distribution at
infinite temperature, where \emph{all} attempts are allowed apart from
those that yield forbidden structures.  
The discrete time and state master equation for the so
constructed chain on the level of the macrostates is given by
\begin{equation}
\label{eq:master-equation}
p(E_j,t+1) = \sum_{E_i} Q(j|i) \cdot p(E_i,t),
\end{equation}
where $p(E_i,t)$ denotes the probability of finding macrostate state
$E_i$ at time $t$ and $Q(j|i)$ is the macrostate transition matrix,
i.e.\ the probability of jumping to a state with energy $E_j$, given
that the random walker sits in state with energy $E_i$.  Since $Q(j|i)$
is stochastic we require that the columns sum to one, i.e.\ $\sum_j
Q(j|i)=1$ for all $i$.  The stationary distribution of
\Eq{eq:master-equation} is the wanted DOS $g(E)$. For a known $Q(j|i)$
this can either  be computed via solving the eigenvalue problem
\Eq{eq:master-equation}, or iterating the equation
\begin{equation} 
\label{eq:stoch-iteration}
p^{(k+1)}(E_j) = \sum_i Q(j|i) \cdot p^{(k)}(E_i)
\end{equation}
starting with arbitrary initial values $p^{(0)}$ \cite{Stroock2004}.

Next we discuss how $Q$ can be estimated from MC data on
attempted moves.  $Q(j|i)$ is related to the microstate transition
matrix $\Gamma(\mathcal{B}|\mathcal{A})$, which is 1 if two non-forbidden
structures are connected by a flip of a pair,
 and to $N(\mathcal{C},\Delta E)$:
\begin{eqnarray}
\label{eq:transition-matrix}
Q(j|i) & = & \frac{1}{C_i} \sum_\mathcal{A} \left( \sum_{\mathcal{B}}
       \Gamma(\mathcal{B}|\mathcal{A}) \cdot
       \delta_{E(\mathcal{B}),E_j} \right) \cdot
       \delta_{E(\mathcal{A}),E_i}\nonumber
       \\ 
       & = & \frac{1}{C_i}
       \sum_\mathcal{A} N(\mathcal{A},E_j - E(\mathcal{A})) \cdot
       \delta_{E(\mathcal{A}),E_i} 
\end{eqnarray}
where $\delta_{i,j}$ is the Kronecker symbol. 
%and $\left< \cdot\right>_E$ denotes microcanoncial average at energy $E$.
  The $C_i$'s are chosen such that $Q(j|i)$ is stochastic.

An sample estimate of $Q(j|i)$ can be made \cite{Anderson1988,Heilmann2005}
from MC data on attempted steps and waiting times: In the matrix
$\hat{W}(j|i)$ we count the number of attempted moves from $i$ to $j$.
Since our approach considers only flips of single pairs, $\hat{W}$
is nonzero only for $j=i-1,i,i+1$. During the simulation 
$\hat{W}(i\pm 1|i)$ is incremented by one for all proposed
transitions (from $i$ to $i\pm 1$),
while $\hat{W}(i|i)$ is always incremented by the waiting time.

Even though the configurations have not to be drawn from a
microcanonical distribution definitely (one might have a canonical
distribution, a mixture of canonical distributions or a generalized
ensemble in mind), we require the samples to fulfill the
{\em microcanonical property}, namely that the probability or weight
of a structure $\mathcal{C}$ on energy $E(\mathcal{C})$ \emph{only}. 
 That means
all states with fixed energy are have equal probability.  That
condition is a crucial point. It was shown that it is automatically
fulfilled, when detailed balance is guaranteed \cite{Wang1999b}.
Simulated annealing violates detailed balance explicitly and therefore
it is hard to prove, that ParQ yields the correct result.  We will
examine this issue in \sect{sec:flatness}.

%%%%%%%%%%%%%%%%%%%%%%%%%%%%%%%%%%%%%%%%%%%%%%%%%%%%%%%%%%%%%%%%%
%
%
%  ParQ 
% 
%
%%%%%%%%%%%%%%%%%%%%%%%%%%%%%%%%%%%%%%%%%%%%%%%%%%%%%%%%%%%%%%%%%
\subsection{ParQ: Estimating the transition matrix }

\begin{figure}
  \begin{center}
    \includegraphics[clip,width=0.75\textwidth,angle=-90]{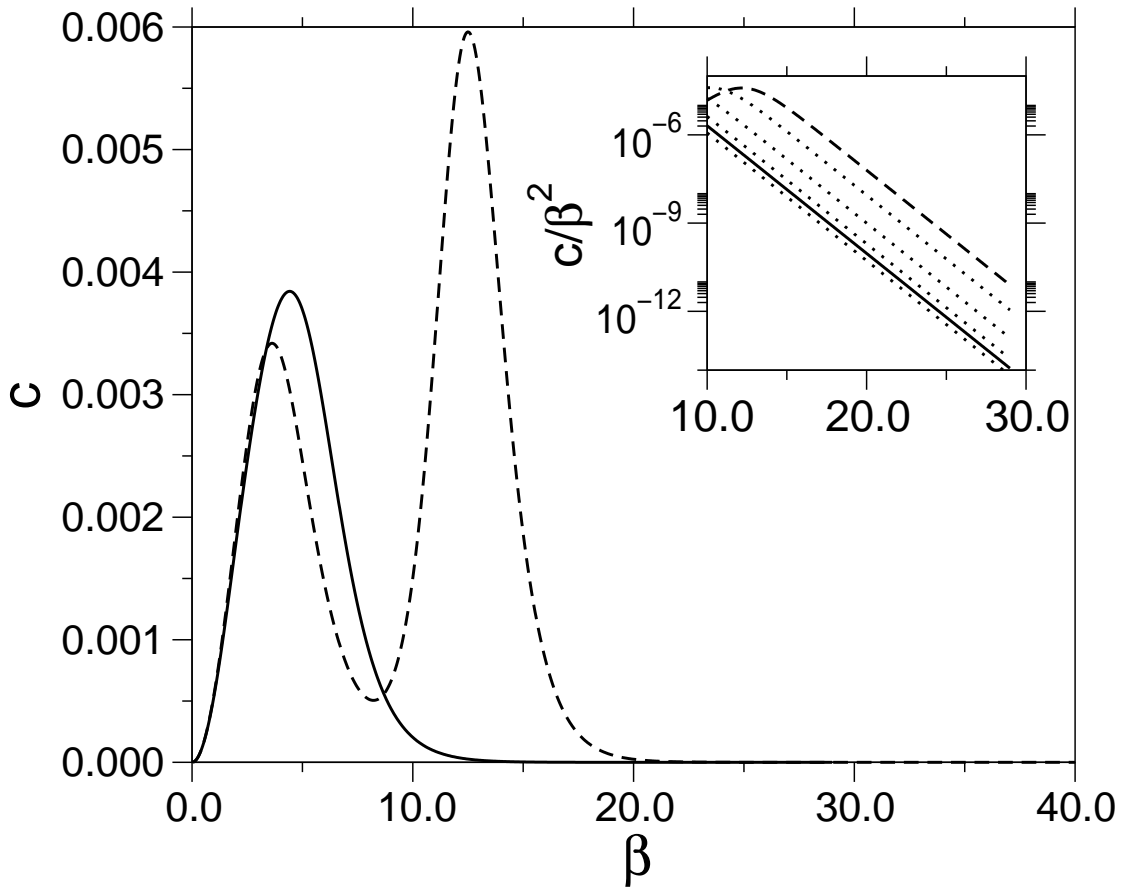}
    \caption{
      \label{fig:HEAT} 
      Specific heat as a function of the temperature, shown for 
      two realization of length $L=80$ with a typical
      (solid line) and a large (dashed) ratio $g(E_1)/g(E_0)$.  Inset:
      low temperature decay rate of the reduced specific heat is the
      same for all realizations $c/\beta^2 \sim \mathrm{e}^{-\beta}$. Dotted
      lines show some other realizations.  }
  \end{center}
\end{figure}

The ParQ \cite{Anderson1988,Heilmann2005} algorithm combines ideas
from simulated annealing \cite{Kirkpatrick1983,Schneider2006} and TMMC.  
Instead of
estimating the transition matrix from an equilibrium simulation the
temperature is lowered according to a certain protocol. The acceptance
rule is the usual Metropolis one
\[
%P_{\text{acc}}(i \rightarrow j) 
r= \min\left(1,\exp [ -\beta \Delta E ]
\right).
\]
%Alex: In (4) hiess die Akzeptanzrate noch r, deshalb habe ich
% es hier auch genommen. P_{acc} wird vorher und nachher nie
% verwendet, also auch hier sinnlos eine neue Größe zu nehmen.

The advantages of the method is first that no assumption about the DOS
is required at the beginning of the simulation.  Secondly, in contrast
to the Wang Landau method, ParQ is easy to parallelize because many
independent runs can be performed simultaneously.

It is required that all regions of interest are visited by the random
walker. Therefore the annealing schedule has to be adjusted. Basically
there are two ingredients: the functional form of the (inverse)
temperature protocol $\beta(t)$ and the start and end value of the
temperature $\beta_1$ and $\beta_2$.
At infinite temperature, the random walker is located at the maximum of
the DOS (see \Fig{fig:DOS}), which corresponds to simple sampling,
where all allowed steps are accepted.  In order to go beyond the
maximum towards the unfolded RNA, i.e.\ for increasing the energy, one has
to chose a \emph{negative} temperature. For the opposite direction,
towards the ground state, the temperature has to be positive and finite. 
 The simplest annealing schedule is
a linear increase of $\beta$ from $\beta_1 < 0$ to $\beta_2 > 0$.
However this kind of cooling schedules might not be optimal.  
Therefore we also checked two other forms, where the inverse temperature
is first increased from $\beta_1$ to a 
certain positive value above the critical temperature, say $\beta=1$ in a 
linear fashion and then
the system is cooled down either linearly or exponentially in
$T$.  We will denote this three schedules as inverse (INV), linear
(LIN) and exponential (EXP) cooling respectively and compare the
performance of the three methods later on.

For infinite long simulations, i.e.\ infinite slow cooling, 
the sampling approaches the
standard Metropolis sampling and the method of estimating the
the transition matrix in this way becomes exact, because
the microcanonical property is fulfilled.  
Unfortunately, the
convergence for a finite number of cooling steps but infinite number
of runs has not been proved so far.

The temperature range $[ \beta_1, \beta_2 ]$ should be chosen, such
that the energy fluctuations vanish sufficiently.  This can be assessed
by considering the specific heat capacity $c = \beta ^2 ( \left< e^2
\right> - \left< e \right>^2)$ \cite{Schneider2006} obtained 
from exact calculations, where
$e$ is the energy per  pair of bases, i.e.\ $e=E/L^2$.  For the usual
case,  where
the DOS is not  a priori $c(\beta)$ known, has to be estimated from a few
primary simulations or the temperature range has be estimated in
other heuristic ways.

The specific heat capacity for two different realizations of length
$L=80$ is shown in \Fig{fig:HEAT}.  For these two examples we used
temperature ranges $[-10,10]$ and $[-10,15]$, respectively.
Note that the decay of $c$ in the low temperature limit $\beta
\rightarrow \infty$ (see inset of \Fig{fig:HEAT}) can be understood very well
\cite{Wang1988}: At low temperatures the partition function is
dominated by the ground state and first excitations only and hence
\[
\frac{c}{\beta^2} = \left< e^2 \right> - \left< e \right> ^2 \sim (E_0
- E_1)^2 \cdot \frac{g(E_1)}{g(E_0)} \cdot \mathrm{e}^{-\beta (E_1-E_0)}
\]
Since $E_1 - E_0 = 1$ for all realization in our simple model the
specific heat capacity decays as $C / \beta^2 \sim \exp( - \beta)$ and
only the prefactor is dominated by large sample to sample fluctuations
of $g(E_1)/g(E_0)$ (see \sect{sect:ratio-stat}).  A large value of
this ratio implies a narrowed peak of the specific heat capacity and
hence increasingly slow relaxation times.  In more complex systems,
such as RNA secondary structure with hybrid energy models
\cite{Burghardt2005}, even the exponent may vary because of variable
energy difference between ground states and first excitations.

%%%%%%%%%%%%%%%%%%%%%%%%%%%%%%%%%%%%%%%%%%%%%%%%%%%%%%%%%%%%%%%%%%%%%%%%%%%%%%%%%%%%%
%
% CONVERGENCE
%
%%%%%%%%%%%%%%%%%%%%%%%%%%%%%%%%%%%%%%%%%%%%%%%%%%%%%%%%%%%%%%%%%%%%%%%%%%%%%%%%%%%%%
\section{Analysis of convergence}
\label{sec:convergence}
In order to assess the performance of different MC algorithms,
we conducted simulations using the different approaches 
described above. We compared the performance using a fixed
realization of length $L=80$ and small ground-state
degeneracy, i.e.\ a large ratio $g(E_1)/g(E_0)$. 
This ratio is somehow a measure for the amount of meta-stable
states. It is a purely local property and does not depend
on large structures of the energy landscape.
Those instances with a large value of this ratio are the
expected to be ``hardest'' instances by comparison with spin
glasses \cite{Adler2004,Dayal2004}, 
and indeed confirmed by our results, see \sect{sec:correlation}.
For all simulations techniques, $5 \times 10^9$ MC steps were used
totally.  We performed various independent simulations (25 for ParQ
and the optimized ensemble and 16 for the flat histogram random
walker) with different seed values but fixed realization.

\begin{figure}
  \begin{center}
    \includegraphics[clip,width=0.97\linewidth,angle=-0]{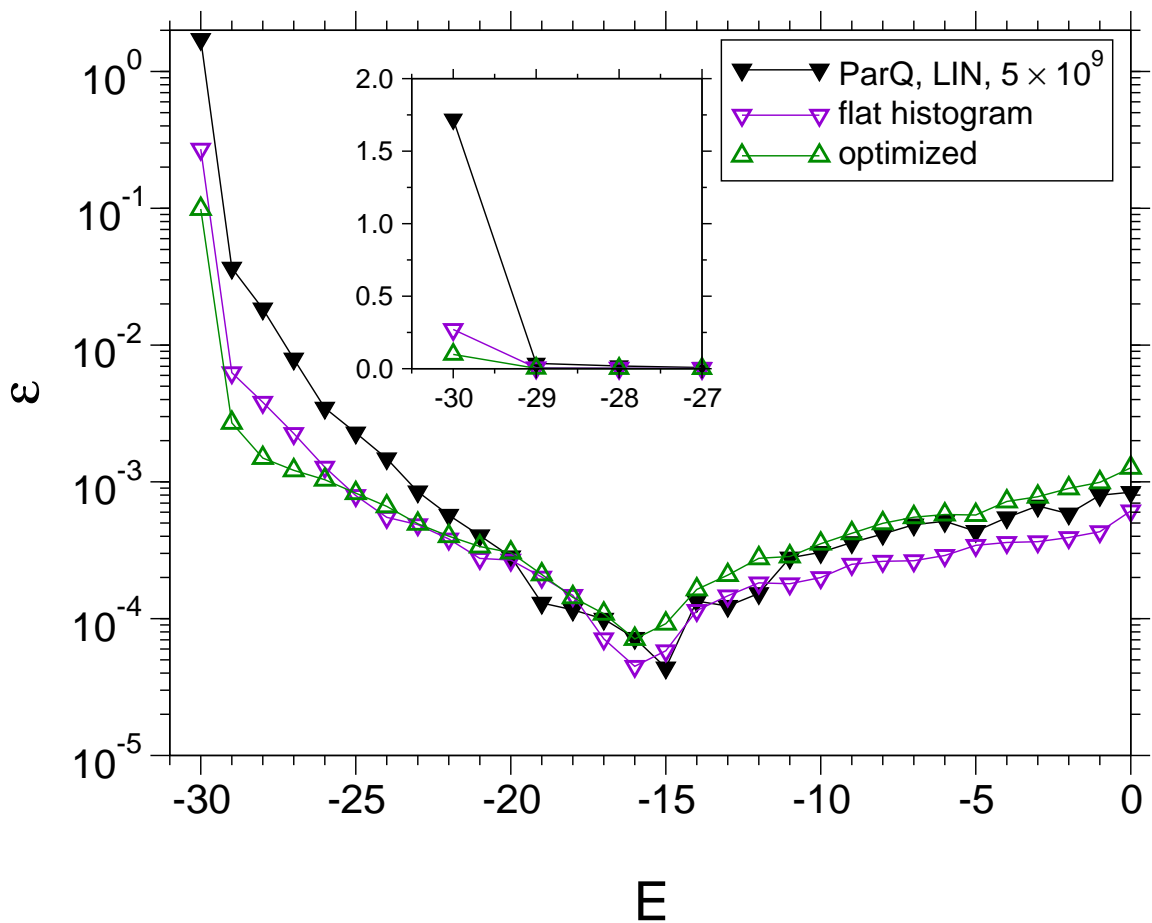}
    \caption{
      \label{fig:DOSERR} 
      Relative error of the DOS of a low degeneracy of ground states
      using flat histogram ensemble, optimized ensemble and ParQ with
      fixed cooling rate ($5 \times 10^{10}$ steps per run).  The inset
      shows the same data with a linear ordinate.  The ratio
      $g(E_1)/g(E_0)$ for the realization was $3120649/16$, which is
      larger than typical values. Lines are guides to the eyes only.  }
  \end{center}
\end{figure}

\begin{figure}
  \begin{center}
    \includegraphics[clip,width=0.97\linewidth,angle=-0]{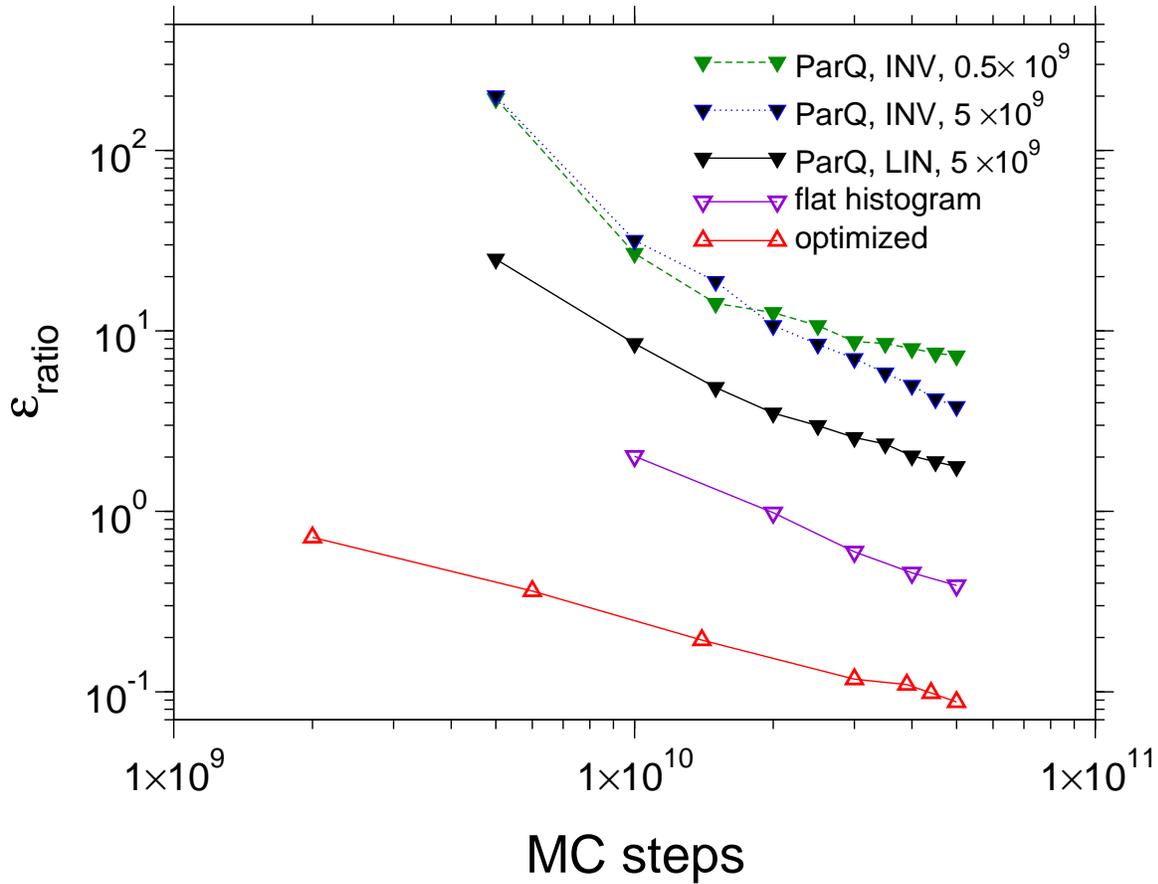}
    \caption{
      \label{fig:CONVRATIO} 
      Rate of convergence of the relative error of the ratio
      $g(E_1)/g(E_0)$ for different simulation methods: ParQ for
      inverse and linear cooling schedule using $500 \times 10^6$ and
      $5000 \times 10^6$ steps per run. For $5000 \times 10^6$ steps
      no significant difference between inverse and linear cooling is
      visible.  Flat histogram and the optimized ensemble sampling
      perform much better than ParQ.  The ratio $g(E_1)/g(E_0)$ for
      the realization was $3120649/16$, which is larger than for typical 
      instances.  }
  \end{center}
\end{figure}

The ParQ result was obtained by a linear and inverse temperature
schedule with a temperature range from $\beta_1= -10$ to $\beta_2=15$
(data for the inverse schedule is not shown in
\Fig{fig:DOSERR}) and the overall $5 \times 10^{10}$ MC steps
were separated in 10 independent runs of length $5 \times 10^9$.  

The so approximated DOS obtained from ParQ
was used as an input for the flat-histogram method, as well as for
the first iteration of
the optimized flat-histogram method. This might be a realistic procedure,
because the DOS is not known in general.  
We used $E_-=E_1$  and $E_+=0$.
For the standard flat-histogram approach, no further adjustment had
to be made, hence the histograms could be sampled using this input
using all available MC steps. For the optimized ensemble,
to optimize the function 
$f(E)$, describing the history of the walker with respect to $E_+$,
we applied $10^9$ MC
steps for the first iteration and then doubled this number always for
each following iteration.
Similar to the $L=40$ system (\Fig{fig:CONVOPT}), the estimate of $f(E)$
converged after only 5 iterations. Hence the optimal
weights where found quickly. Via this optimization, the round-trip time
decreased by a factor of about $4$. 
The remaining $1.9\times
10^{10}$ steps, where the weights were kept fixed, could be used
to gather the final histogram.

To compare the power of the different algorithms,
we considered the relative error of the MC approximation with
respect to the exact solution $\epsilon(E) = \left| \hat{g}(E) - g(E)
\right| / g(E)$, where $\hat{g}$ is the sample estimate obtained 
either by iteration of \Eq{eq:stoch-iteration} or via the
energy histogram, depending on the algorithm.  
The averaged $\epsilon(E)$ is shown in \Fig{fig:DOSERR}.  

A second quantity, which gives a relevant measure of performance is
the sample error of ratio between first excitations and ground states
\[
\epsilon_{\text{ratio}} = \frac{ \left| \hat{g}(E_1) / \hat{g}(E_0) -
g(E_1) / g(E_0) \right| }{ g(E_1) / g(E_0) }.
\]
This quantity as a function of MC steps is shown in \Fig{fig:CONVRATIO}.

>From \fig{fig:DOSERR} and \fig{fig:CONVRATIO} one can learn that in
the high energy region were only a few sites are connected by bonds
the flat histogram method clearly outperforms the other methods,
whereas in the relevant low energy region the optimized random walker
seems to be best. The most significant difference between the methods
is located at the ground state of the system, where the ParQ method is
not very accurate.  Also the rate and form of the annealing schedule
affects the performance: The linear schedule seems to outperform the
inverse schedule and, as expected, few long runs beat many short ones.

Note that \fig{fig:DOSERR} and \fig{fig:CONVRATIO} are worst case
scenarios, because we picked out a sample, where the ratio
$g(E_1)/g(E_0)$ is very large, i.e.\ there are many meta-stable states
that might be separated by large barriers from ground states. We also
performed the same kind of simulations for a typical realization of
the same length, where $g(E_1)/g(E_0)$ is relatively small. The
errors of the ratio decrease by a factor of 9.5, 35 and 39 for the
ParQ, flat histogram and optimized ensemble method respectively.

In order to check, if the qualitative ranking of the methods, i.e.
$\epsilon_{\text{ratio}}^{\text{optimized}} <
\epsilon_{\text{ratio}}^{\text{flat}} <
\epsilon_{\text{ratio}}^{\text{ParQ,LIN}}$, is a general feature of the
system we generated an ensemble of 2000 realizations of length $L=40$
and performed the same kind of simulations as before with $5 \times
10^7$ steps for all simulations.  In the majority of the cases (59\%)
we find the same kind of ranking and second most frequently (33\%) a
ranking of $\epsilon_{\text{ratio}}^{\text{flat}} <
\epsilon_{\text{ratio}}^{\text{optimized}} <
\epsilon_{\text{ratio}}^{\text{ParQ,LIN}}$.  Only in $2\%$ percent of
the cases ParQ outperforms one of the generalized ensemble methods.  
Sample averages of $\epsilon_{\text{ratio}}^{\text{optimized}}$,
$\epsilon_{\text{ratio}}^{\text{flat}}$ and
$\epsilon_{\text{ratio}}^{\text{ParQ,LIN}}$ were $0.030$, $0.055$ and
$0.551$ respectively. Probably these differences  increase for larger
systems.

We also checked that linear cooling is better than the other two
alternatives in $53\%$ of the cases (exponential and inverse cooling
only $15\%$ and $31\%$ respectively).

%%%%%%%%%%%%%%%%%%%%%%%%%%%%%%%%%%%%%%%%%%%%%%%%%%%%%%%%%%%%%%%%%%%%%%%%%%%%%%%
%
% COMPLEXITY MEASURES
%
%%%%%%%%%%%%%%%%%%%%%%%%%%%%%%%%%%%%%%%%%%%%%%%%%%%%%%%%%%%%%%%%%%%%%%%%%%%%%%%

\section{Correlation between algorithmic and structural complexity}
\label{sec:correlation}
As already mentioned the performance strongly depends on the ratio
$g(E_1)/g(E_0)$, which was also obtained in $\pm J$ spin glasses
\cite{Adler2004,Dayal2004}. 
In this section we want study  the distribution
of this ratio and its relationship to the performance of MC
algorithms.  We also check if there is a correlation between the
degree of ultrametricity at finite temperature and performance.

\subsection{Ratio between number of first excitations and groundstates}
\label{sect:ratio-stat}

For the usual 2d $L \times L$ Ising ferro magnet without disorder it
is obvious, that the ratio $g(E_1)/g(E_0)$ scales as $L^{2}$, 
because there are
exactly $L \times L$ possibilities to excite the ground state by one
single spin flip.  In our model, RNA secondary structure, the scaling
behavior can not be obtained with such simple arguments.

Hence, we generated ensembles of up to $40,000$ realizations for sequence
lengths between $L=20$ and $L=1021$ sampled the distribution of
$g(E_1)/g(E_0)$.  Even though the transfer matrix algorithm is
polynomial, the computations of systems larger than $L=320$ become very
time consuming.  Therefore we only computed the number of ground states and
first excitations instead of the complete energy spectrum.  This can
be achieved by truncations of the polynomials in the transfer matrix
after the term of the second largest power.

\begin{figure}
  \begin{center}
    \includegraphics[width=0.98\linewidth,angle=0]{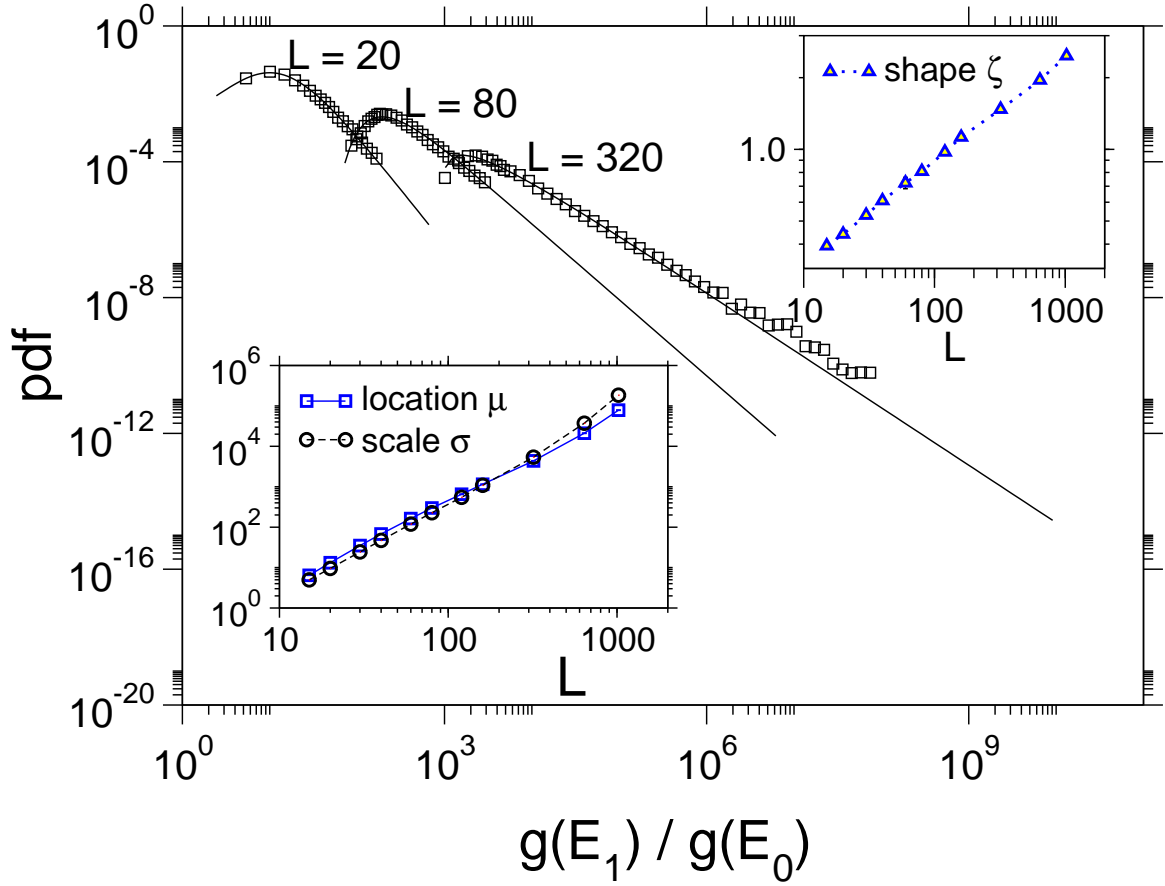}
    \caption{
      \label{fig:RATIODISTR} 
      Density distribution of the ratio $g(E_1)/g(E_0)$ for different
      sequence lengths. Squares indicate binned data.  The largest
      errorbar is as large as the symbols.\\ Insets: scaling of the
      location, scale and shape parameter on a double logarithmic scale.
      }
  \end{center}
\end{figure}

Empirically we find a generalized extreme-value distribution (see
\fig{fig:RATIODISTR} ), whose cumulative distribution function is
given by
\begin{equation}
\label{eq:gev}
\Prob\left(\frac{g(E_1)}{g(E_0)} \geq x\right) = \exp \left[ - \left ( 1+\xi\,\frac{x -
\mu }{\sigma} \right) ^{-1/ \xi } \right],
\end{equation} 
similar as in \cite{Dayal2004}.

The parameters of the distribution $\mu$ ({\em location}), $\xi$ ({\em
shape}) and $\sigma$ ({\em scale}), were obtained through a maximum
likelihood fit using the {\tt FORTRAN} program by Hosking
\cite{Hosking1985,Macleod1989}.  The resulting probability density
functions (pdf) and the scaling behavior of the fit parameters are
shown in \fig{fig:RATIODISTR}.

The qualities of the maximum likelihood fits were not good enough to
be convinced that the data indeed follows \Eq{eq:gev}, especially for large
sequences.  This is also supported by small p-values of
Kolmogorov-Smirnov tests.  But the data at least allows to distinguish
between an exponential and algebraic growth of the location and scale
parameter:
Similar as for the $\pm J$ model \cite{Saul1994} we find an algebraic
behavior of location and shape parameter $\mu(L) = A \cdot L^{z_\mu}$
and $\xi(L) = B \cdot L^{z_\xi}$ with an exponents of $z_\mu =
2.1(1)$ and $z_\xi = 2.4(9)$.  Although the quality of the fit is not
very high (as can be seen already in the lower left inset of
\Fig{fig:RATIODISTR} where the empirical data do not follow a straight line in
the log-log plot), an exponential scaling can be safely excluded
by our data.

%%%%%%%%%%%%%%%%%%%%%%%%%%%%%%%%%%%%%%%%%%%%%%%%%%%%%%%%%%%%%%%%%%%%%%%%%%%%%%%%%%%%%%%%%%%%%%%%%%%%%%%%%%%%%
%
% Ultrametricity 
%
%
%%%%%%%%%%%%%%%%%%%%%%%%%%%%%%%%%%%%%%%%%%%%%%%%%%%%%%%%%%%%%%%%%%%%%%%%%%%%%%%%%%%%%%%%%%%%%%%%%%%%%%%%%%%%%
\subsection{Ultrametricity of the phase space}
\label{sect:ultra}
The study of ultrametric spaces dates back many decades and has entered the
physical literature in the context of spin glass theory (see
\cite{Rammal1986} and references therein).

Higgs found evidence that RNA secondary structures exhibit an ultrametric
structure \cite{Higgs1996} at low temperatures. The existence of a
phase transition  was  then confirmed
numerically by Pagnani et al.\ \cite{Pagnani2000} by considering the width of the 
overlap distribution 
and then examined by other authors using 
droplet theory \cite{Bundschuh2002}, the $\epsilon$-coupling method \cite{Krzakala2002} and 
renormalized field theory \cite{Lassig2006}.

Ultrametricity
can be detected by considering the ``distance'' between two
structures drawn from a canonical ensemble at a given
temperature. Using the transfer matrix $Z_{i,j}$ it is possible to
draw states directly without performing Markov
chain MC \cite{Higgs1996}.  It is also possible to enumerate
ground states or, if the degeneracy is too large to enumerate, one may 
sample them equally likely \cite{Hartmann2001}.

Given two structures $\mathcal{A}$ and $\mathcal{B}$, their overlap is
defined by $q_{ \mathcal{A} \mathcal{B}} = \frac{1}{L} \sum_{n=1}^L
\delta_{ {b^{\mathcal{A}}_n},b_n^{\mathcal{B}}}$, where
\[
  b^{\mathcal{S}}_i = 
  \cases{
    j & { if } $(i,j) \in \mathcal{S}$\\ 
    0 & { otherwise}
    }
\]
The normalized distance between $\mathcal{A}$ and $\mathcal{B}$ is
given by $d(\mathcal{A},\mathcal{B}) = 1-q_{ \mathcal{A}
\mathcal{B}}$.

An ultrametric space $M$ is defined by following axioms:
\begin{enumerate}[(i)]
\item{ $0\leq d(\mathcal{A},\mathcal{B}) $ and $ d(\mathcal{A},\mathcal{B}) = 0 \iff%\leftrightarrow 
\mathcal{A}=\mathcal{B}$}
\item{ $ d(\mathcal{A},\mathcal{B}) = d(\mathcal{B},\mathcal{A})$ }
\item{ $ d(\mathcal{A},\mathcal{C}) \leq \max \left\{
d(\mathcal{A},\mathcal{B}), d(\mathcal{B},\mathcal{C}) \right\}$ },
\end{enumerate}
for all $\mathcal{A}$, $\mathcal{B}$, $\mathcal{C} \in M$.  
Note that every ultrametric space is a metric space.

A direct consequence is that each triangle is isosceles, i.e.\ the two
largest sides of a triangle are of equal length.  This property provides a
numerical criterion for the detection of ultrametricity
\cite{Higgs1996}: The degree of ultrametricity can be estimated by
the difference of the two largest distances of a set of candidate
triangles. This quantity, which is denoted as $s$, vanishes in
perfectly ultrametric spaces and become small in approximately
ultrametric spaces.  

There might be different reasons, why $s$ may
vanish in the thermodynamic limit and therefore one has to filter real
ultrametricity against trivial one. E.g.\ in the high temperature phase, $s$
also scales as $N^{-1/2}$ and the maximum size of $s$ is limited by
the triangle equation. To distinguish from this trivial ultrametricity,
Higgs  samples \cite{Higgs1996}
sets of ``uncorrelated'' triangles by computation of three independent
distances $d(\mathcal{A},\mathcal{B}), d(\mathcal{C},\mathcal{D}),
d(\mathcal{E},\mathcal{F})$.  
If these distances fulfill the triangle
inequalities, i.e.\ $d(\mathcal{A},\mathcal{B}) 
\leq d(\mathcal{C},\mathcal{D}) + d(\mathcal{E},\mathcal{F})$ 
and for all other combinations of the distances,
the difference between the two largest distances is computed.
Finally the average of the differences taken over all 
valid uncorrelated triangles $s_{\text{uncor}}$ is computed.
Non-trivial ultrametricity should emerge faster than the
trivial ultrametricity obtained from uncorrelated distances.
Hence $s^\ast=s/s_{\text{uncor}}$ should
vanish in the presence of an ultrametric structure in the thermodynamic
limit.

\begin{figure}
  \begin{center}
    \includegraphics[clip,angle=-90,width=0.95\linewidth]{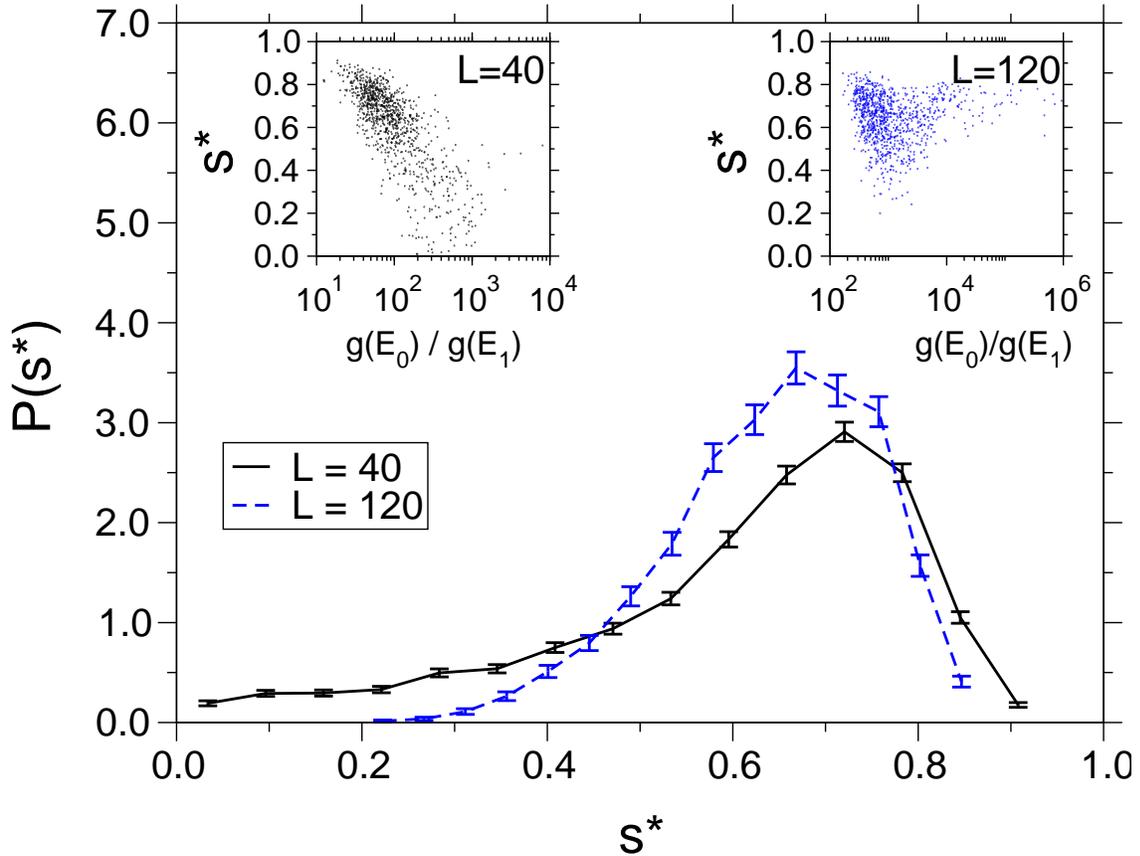}
    \caption{
      \label{fig:ULTRADISTR} 
      Distribution of deviations from ultrametricity at $T=0.125$.
      Inset: scatter plot between $g(E_1)/g(E_0)$ and $s^\ast$.  }
  \end{center}
\end{figure}

In principle one should distinguish the finite temperature and zero temperature behavior 
in complex phase, as already pointed out in \cite{Hartmann2001}. 
Using direct sampling of ground states a ``non-trivial'' overlap distribution at zero temperature could be 
ruled out by numerical extrapolation. This implies that grounds states alone are \emph{not ultrametric}. 
For this reason we considered only finite temperature
states, where the overlap distribution is non-trivial \cite{Pagnani2000} and evidence for an ultrametric phase 
space still remains. This arguments are also supported by the observation, that the temperature dependent values 
of $s^\ast$ have a minimum below the critical and above zero temperature. 
However $s^\ast$ decreases only slowly with system size \cite{Higgs1996} and therefore the quality of 
the data have to be used with  care (see for example \cite{Contucci2007} an the related comment
\cite{Jorg2007}). 
Hence, much larger systems sizes, which are unfortunately out of reach,
 would be necessary to decide this question finally.
However, since our model has properties of a mean field model, an 
ultrametric space would be not surprising (in contrast to 
the finite dimensional model considered in \cite{Contucci2007,Jorg2007}).

In \Fig{fig:ULTRADISTR} we illustrate the distribution of $s^\ast$ 
over many realizations for $L=40$ and $L=120$ at a temperature
$T=0.125$ slightly below the
transition. As expected, the transition moves a bit to towards
small values of $s^\ast$ for increasing system size. 
In the small system the 
correlations between the ratio
$g(E_1)/g(E_0)$ and $s^\ast$ (see insets) 
are stronger. We assume that this is an
finite-size effect, because this effect is weaker for larger systems.  

We will quantify these correlations in the following section.

\begin{figure}
  \begin{center}
    \includegraphics[clip,width=0.50\linewidth]{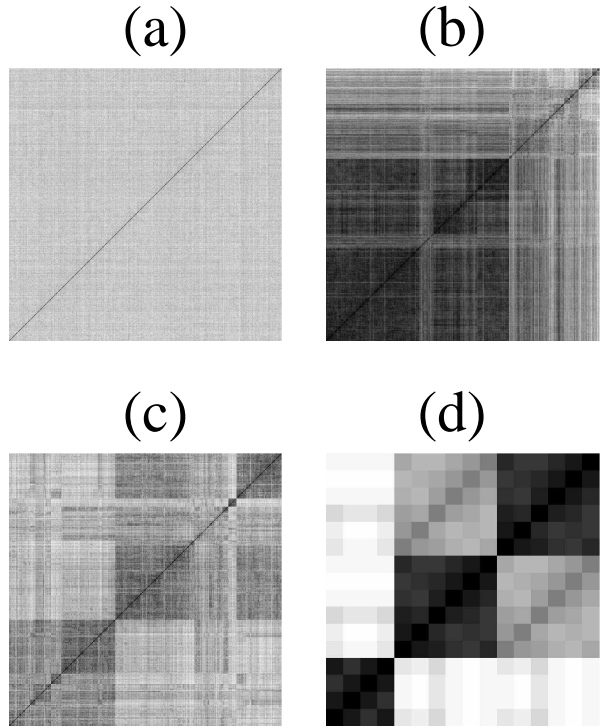}
    \caption{
      \label{fig:DISTMATRIX} 
      Hierarchical structure of the states illustrated by distance
      matrices.  Darker grey scales correspond to large overlaps.  (a)
      $L=120$ at $T=2$. No hierarchical structure could be detected 
($s^\ast\approx 1$)\\
      (b) $L=120$ at $T=0.125$ for a realization exhibiting a weak
      ultrametricity ($s^\ast \approx 0.74$)\\ (c) $L=120$ at
      $T=0.125$ for a realization exhibiting stronger ultrametricity ($s^\ast
      \approx 0.45$)\\ (d) $L=40$ at $T=0.0$ for a realization having low
      ground-state degeneracy ($g(E_1)/g(E_0) = 14638/16$) Deviation
      from ultrametricity was $s^\ast \approx 0.5$.  Realization (d)
      was also used in section \ref{sec:flatness}. The corresponding
      dendrogramm is illustrated in \Fig{fig:HIST}.  }
  \end{center}
\end{figure}

But first we explain an alternative approach, where
ultrametric spaces can also be detected and 
analyzed by clustering low distance
configurations in hierarchical groups.  The clustering algorithm used
in this paper is Ward's algorithm \cite{JainDubes1988}, an
agglomerative hierarchical matrix updating algorithm, also called
minimum variance method as it is designed to minimize the variance of
the constructed clusters.  The algorithm works as follows. Initially
each configuration forms a cluster of its own, and the distance matrix
$d_{ab}$ with the distances of all pairs of clusters(each containing
one configuration) is calculated using e.~g. the above defined
distance.  Then in each step the two clusters $p$ and $q$ with the
smallest distance are fused to form a new cluster $t$.  The distance
matrix is updated using
\begin{equation}
  \label{eq:ward}
   d_{rt} = \frac{(n_r+n_p)d_{rp}+(n_r+n_q)d_{rq}-n_rd_{pq}}{n_r+n_t}
\end{equation}
where $r$ refers to any of the other clusters left unchanged in the
current step and $n_x$ is the number of configurations in cluster $x$.
This is repeated until only one cluster remains which now contains all
configurations.  Afterward, one can re-order the configurations
according to the cluster hierarchy obtained in the fusing process, and
draw a color-coded visualization of the distance matrix.  In
\fig{fig:DISTMATRIX} four different distance matrices for different
realizations and temperatures are illustrated. As one can see, a clear
cluster structure emerges only at low temperatures, but a clear
correspondence between the ultrametricity $s^\ast$ and the visual
impression cannot be made.
Note that one can apply a clustering algorithm to any set of data, hence
also to non-ultrametric ones. There are quantitative methods, which
test how much the tree imposed by the clustering algorithm correlates
with the distances in the data. Here, we have just used the visual
impressions obtained by looking at the matrices. Furthermore,
in the last section we will use the so detected clusters to check
whether all ground states are visited with equal probability.

%%%%%%%%%%%%%%%%%%%%%%%%%%%%%%%%%%%%%%%%%%%%%%%%%%%%%%%%%%%%%%%%%%%%%%%%%%%%%%%%%%%
%
% Tunneling time distribution.
%
%%%%%%%%%%%%%%%%%%%%%%%%%%%%%%%%%%%%%%%%%%%%%%%%%%%%%%%%%%%%%%%%%%%%%%%%%%%%%%%%%%%%
\subsection{Distribution of tunneling times of the flat histogram random walker}

We measured the tunneling time for the flat histogram random walker
for sequence lengths up to $120$. Note that larger systems become
infeasible if one wants to span a large energy interval, which we need
to when studying tunneling time distributions. Already for $L=120$ we found 
tunneling times fluctuating ranging in real time 
between seconds and days on a modern CPU.

\begin{figure}
  \begin{center}
    \includegraphics[clip,width=0.75\textwidth,angle=-90]{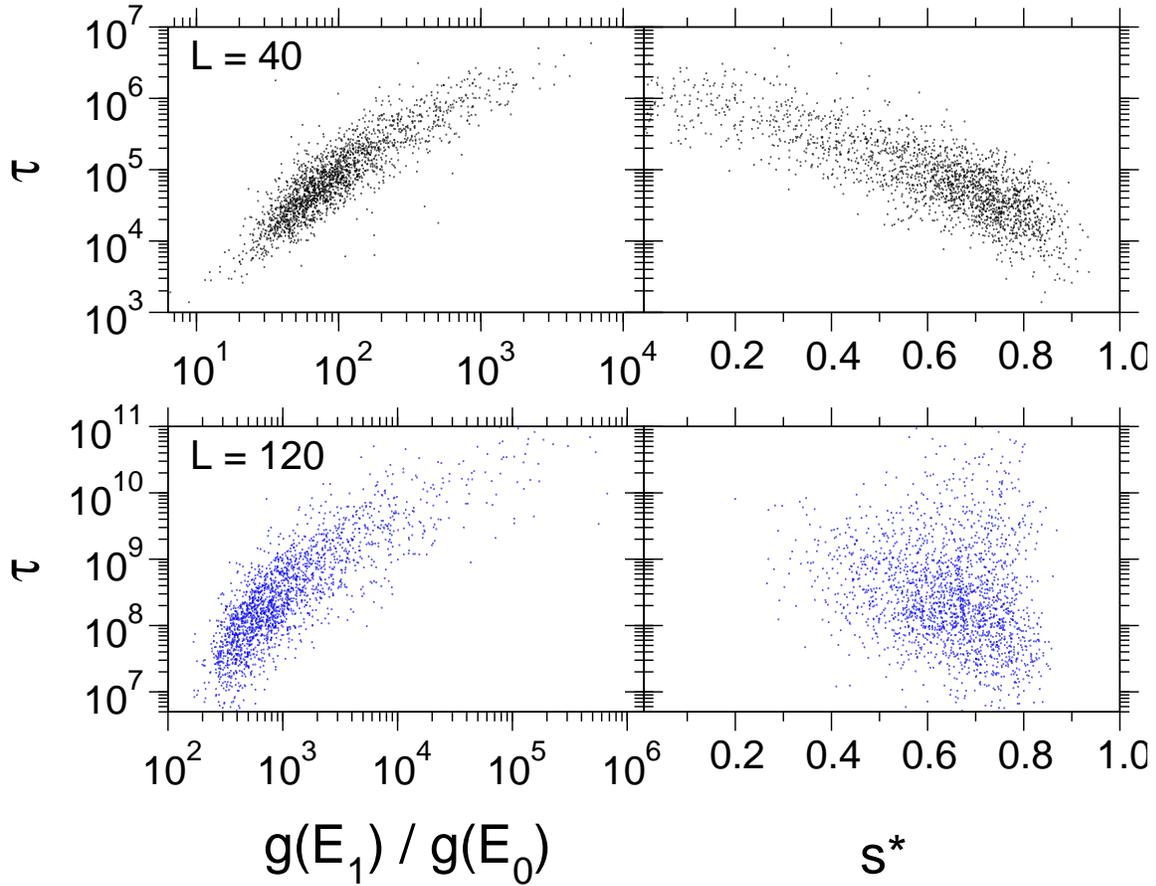}
    \caption{
      \label{fig:TUNNELCORR} 
      Left column: Scatter plot between the ratio $g(E_1)/g(E_0)$ and
      tunneling time of the flat histogram sampler for $L=40$ (upper)
      and $L=120$ (lower). \\ Right column: Scatter plot between
      deviation from ultrametricity and  tunneling time of the flat
      histogram sampler for $L=40$ (upper) and $L=120$ (lower).  }
  \end{center}
\end{figure}

There is a strong correlation between $g(E_1)/g(E_0)$ and the tunneling time,
see \Fig{fig:TUNNELCORR}. We found that this is in particular true
when this ratio is much larger than
all other ratios between neighboring energy densities.  The
performance of the algorithm is then dominated by the rare event of
finding a ground state when starting 
from a first excitation.  Two scatter plots of the
ultrametricity index $s^\ast$ versus tunneling time are shown in
\fig{fig:TUNNELCORR}. Hence, whether there
is a true correlation between tunneling time $\tau$ and degree of
ultrametricity $s^\ast$ is not clear, because for larger system
the correlation appears to be rather weak.

To investigate the issue of correlations between static
measures and computational hardness more quantitatively we
calculated the empirical Pearson
correlation coefficients for all pairs of  quantities $\log \tau$, $\Delta S =
\log \left(g(E_1)/g(E_0)\right)$ and $s^\ast$, which results in
\[
\hat{\rho} = \left(
\begin{array}{ccccccc}
  & \vline & \log \tau & \Delta S & s^\ast_{(T=0.125)} &  s^\ast_{(T=0.033)} \\ 
  \hline \log \tau & \vline & 1 & 0.89 & -0.40 &  -0.33 \\ 
  \Delta S & \vline & & 1 & -0.37 & -0.30 \\
  s^\ast_{(T=0.125)}& \vline & & & 1 & 0.87 \\ 
  s^\ast_{(T=0.033)}&  \vline & & & & 1
\end{array} \right)
\]
for $L=40$ and
\[
\hat{\rho} = \left(
\begin{array}{ccccccc}
  & \vline & \log \tau & \Delta S & s^\ast_{(T=0.125)} & s^\ast_{(T=0.033)} \\ 
  \hline \log \tau & \vline  & 1 & 0.82 & -0.18 & -0.16 \\ 
  \Delta S & \vline & & 1 & 0.02 & -0.13 \\ 
  s^\ast_{(T=0.125)}& \vline & & & 1 & 0.28 \\ 
  s^\ast_{(T=0.033)}&  \vline & & & & 1 
\end{array} \right)
\]
for $L=120$.  As pointed out in \sect{sect:ultra} there is a
correlation between $s^\ast$ and $\Delta S$, at least for smaller
systems.  In order to check if there is a direct correlation between
ultrametricity and tunneling time we computed the partial correlation
$\hat{\rho}_{\log \tau, s^\ast \cdot \Delta S}$ \cite{Afifi}, 
which leads to no
significant values for all sequence lengths we considered here. Hence
this correlation might be trivial, because it is induced by
correlation of the ratio $g(E_1)/g(E_0)$ and tunneling time. 

This means, although ultrametricity is usually considered as a
landmark of complex and glassy systems, at least for the behavior of
RNA secondary structures it is not related to the dynamic glassy behavior
seen in MC simulations. 
We believe that the effect of ultrametricity is superimposed by 
the large number of metastable states. 

\begin{figure}
  \begin{center}
    \includegraphics[clip,width=0.95\linewidth,angle=-00]{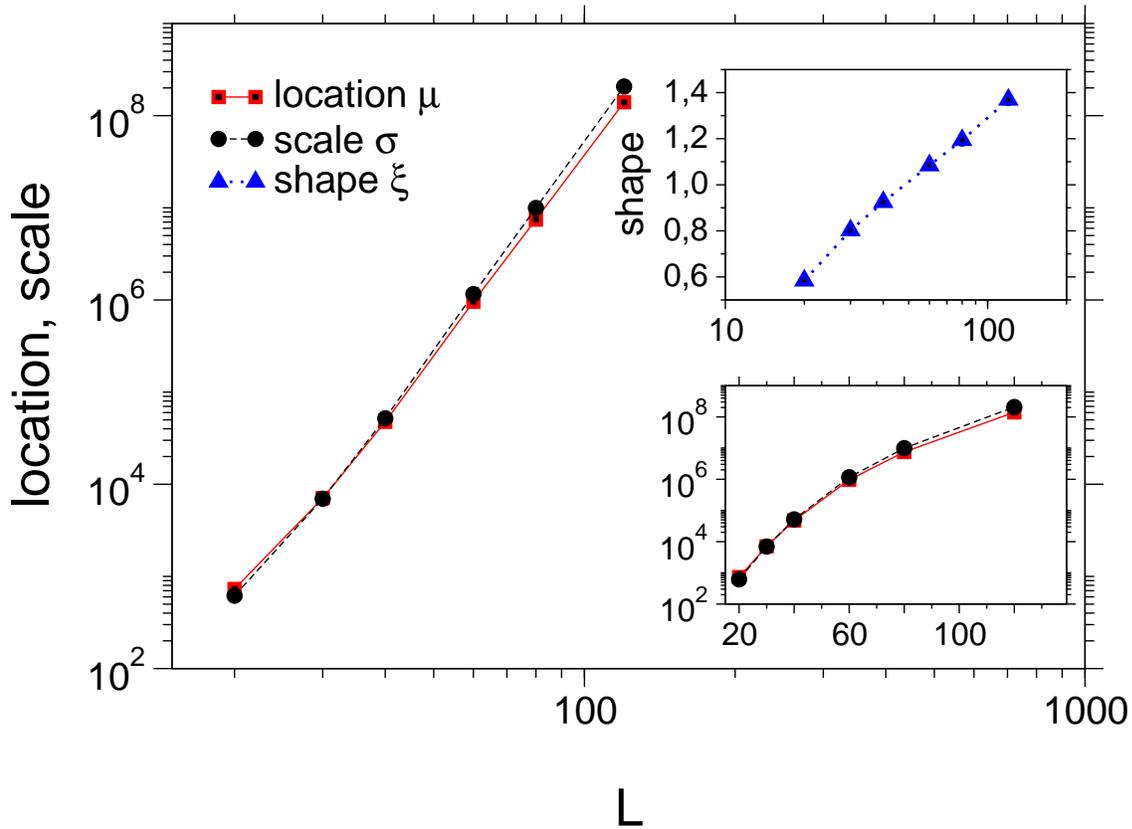}
    \caption{
      \label{fig:TUNNELSCALE} 
      Scaling of the location (square symbols) 
     and scale parameters (circles) on the tunneling
      time.  Insets: scaling of the shape parameter using a logarithmic
      abscissa (top) and scaling of the location and scale parameters
      using a logarithmic ordinate (bottom).  }
  \end{center}
\end{figure}

We fitted also the distributions of the tunneling time
to a generalized extreme-value distribution \Eq{eq:gev}
and analyzed the scaling of the
parameters. 
%Location and scale parameter are always almost equal 
%depend algebraically on the sequence length, see \Fig{fig:TUNNELSCALE}.
Location and scale parameter have almost the same algebraic dependence on the
sequence length, see \Fig{fig:TUNNELSCALE}. 
As can be seen on the semi-logarithmic plot in the bottom inset in
\Fig{fig:TUNNELSCALE} an exponential scaling can safely be excluded
(at least on the length scale up to $120$).  The exponent describing the
power-law is roughly $z \approx 7$. On the other hand,
the shape parameter seems to scale logarithmically in
sequence length, see upper inset in \Fig{fig:TUNNELSCALE}.
 With the same arguments as in
\sect{sect:ratio-stat}, we cannot exclude that the distribution deviates 
from a generalized extreme-value distribution. 

These results differ from the $\pm J$ Ising model, where an exponential
tunneling time was observed. On the other hand, they
are  similar to the fully frustrated
Ising model, investigated by Dayal et al.\ \cite{Dayal2004}  They
argued, that sub-exponential growth of tunneling
times and of the  ratio $g(E_1)/g(E_0)$ 
 stem from a smaller growth of the number of meta-stable states.
These results suggest that the model of RNA is 
dynamically ``simpler'' than $\pm J$ spin glasses and has 
a similar complexity as the fully frustrated model.

On the other hand sample to sample fluctuations are much 
larger than in the $\pm J$ model, as can be seen by 
comparing the shape parameter in the range of investigated system sizes.
For the largest systems in \cite{Dayal2004}, the scaling
parameter was about $0.9$. Hence, although typically RNA instances are
not so hard for an MC algorithm, compared to $\pm J$ spin glasses,
there is larger fraction of rare hard instances for RNA secondary
structures.

\subsection{Distribution of MC errors of ParQ and the 
optimized ensemble random walker} 
\begin{figure}
  \begin{center}
    \includegraphics[clip,width=0.75\textwidth,angle=-90]{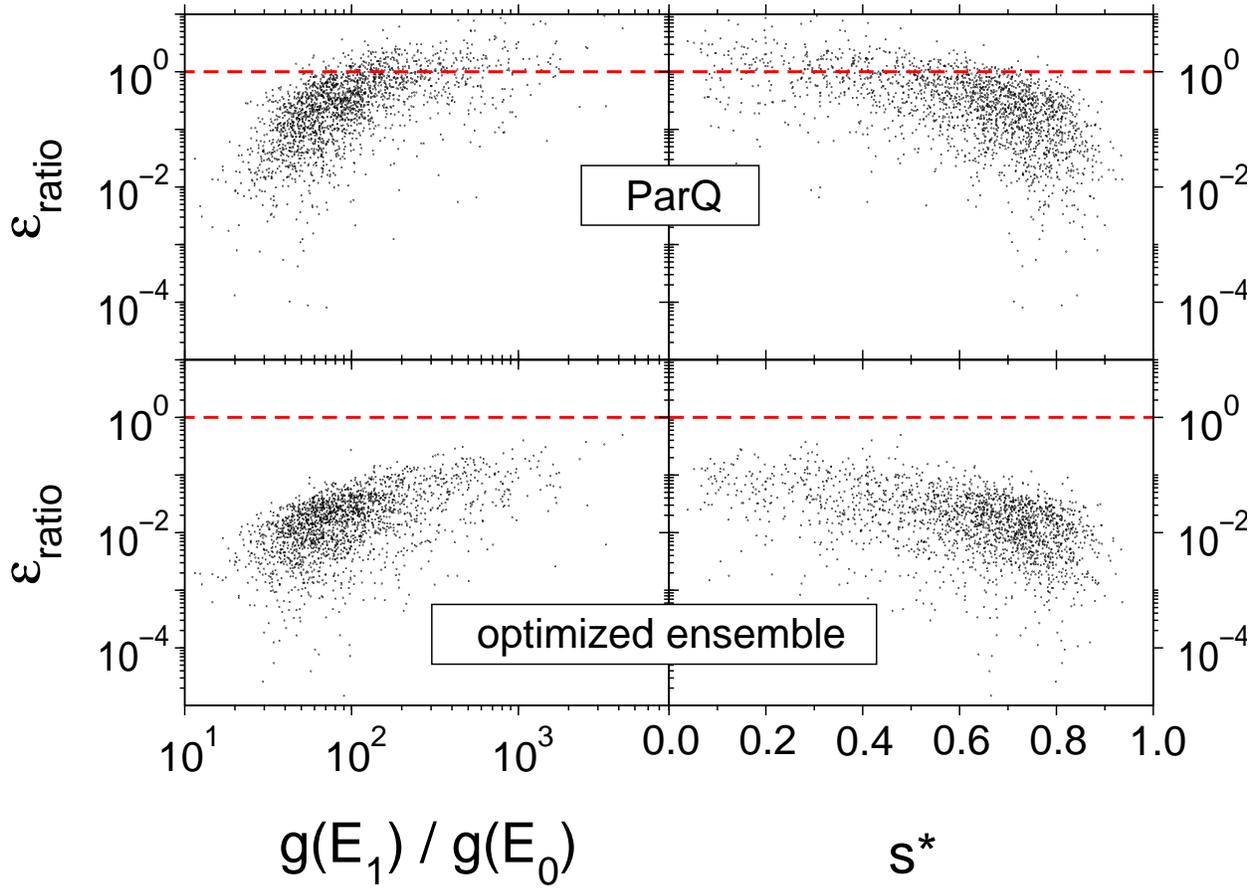}
    \caption{
      \label{fig:ERRORCORR} 
      Correlation between the ratio $g(E_1)/g(E_0)$ and the error of
      the ratio for ParQ using linear cooling (upper left) and
      optimized ensemble simulation (lower left) as well the
      correlation between deviation from ultrametricity
      $s/s_{\text{uncor}}$ and the error (upper right for ParQ and
      lower right for optimized ensemble).  The error for optimized
      ensemble is about one order of magnitude smaller than for the
      optimized ensemble simulation.  }
  \end{center}
\end{figure}

Next the correlation between sampling errors of the ratio $g(E_1)/g(E_0)$, the
value of the ratio itself, and the degree of ultrametricity was considered
(see \Fig{fig:ERRORCORR}).  We used the ensemble of realizations of
length $L=40$, which was already used in \sect{sec:convergence}.

Again we find a correlation between the ratio and performance, which
can be quantified via correlation coefficients.  The correlation between
the error $\epsilon$ and the log-ration $\Delta S$ was larger
using the ParQ method 
($\hat{\rho}(\epsilon^\text{ParQ}_\text{ratio}, \Delta S) \approx 0.6$)
compared to the correlation for the optimized ensemble
($\hat{\rho}(\epsilon^\text{optimized}_\text{ratio}, \Delta S)
\approx 0.5)$.  Similar to the preceding section,
a significant partial correlation between degree of
ultrametricity and sampling errors for fixed log ratio $\Delta S$
could not be detected.

\subsection{Are all ground states visited with equal probability?}
\label{sec:flatness}

\begin{figure}
  \begin{center}
    \includegraphics[clip,width=0.75\textwidth,angle=-90]{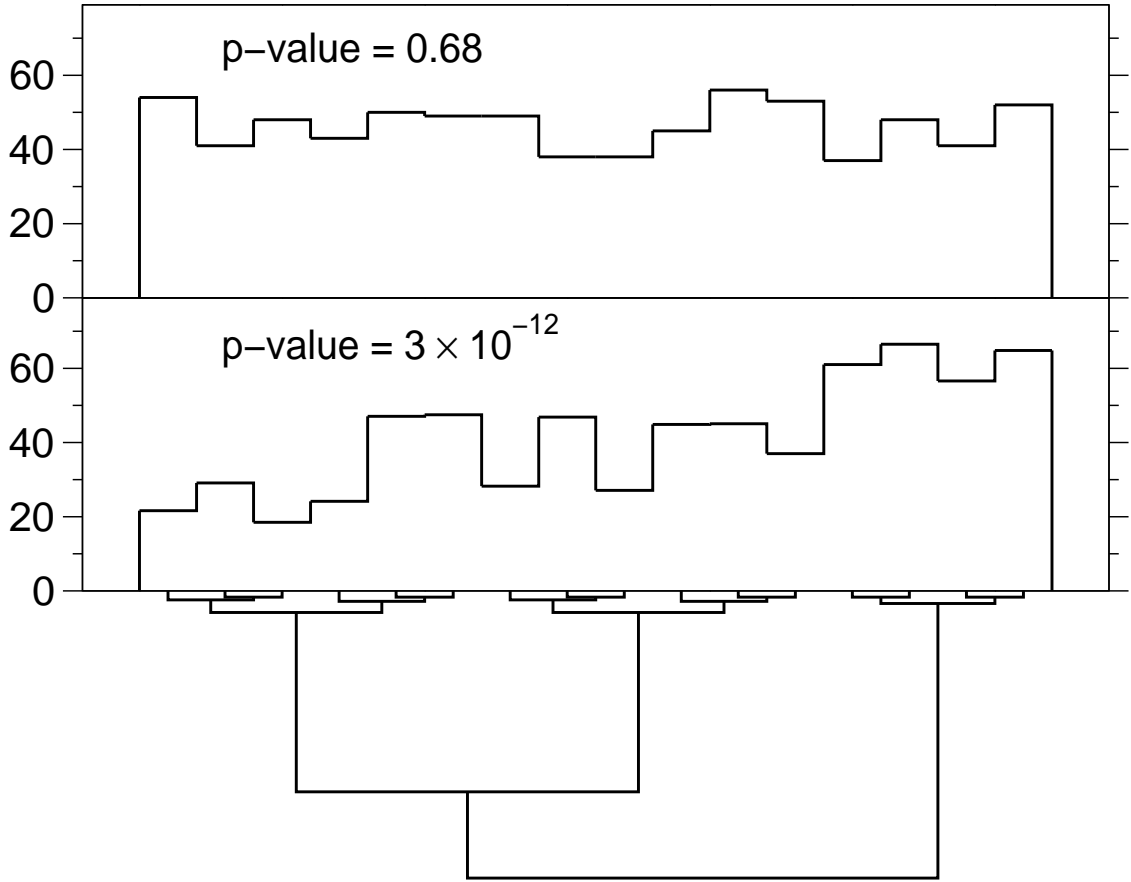}
    \caption{
      \label{fig:HIST} 
      Histograms of visited ground states. upper: optimized ensemble
      random walker, lower simulated annealing.  The Bhattacharyya
      measures and the corresponding p-values indicate, that simulated
      annealing does not visit all ground states with equal weight.  }
  \end{center}
\end{figure}

>From \Fig{fig:ERRORCORR} one can also see that the error for the
optimized ensemble are one order of magnitude smaller than that of
ParQ. Since in both cases the data were obtained from the transition
matrix the significant difference must be caused in the underlying MC
scheme, probably the non-equilibrium character of ParQ. In order to
gain insight to this issue we checked whether the microcanonical property
is fulfilled.
%In detail we checked if all microstates for a fixed energy are visited with equal probability 

In detail we considered histograms of visited ground states for
simulated annealing (ParQ) and the optimized ensemble sampler and
checked if the histograms were sufficiently flat.
A simple and powerful check for the flatness of a histogram is the
Bhattacharyya distance measure (BDM) for two given probability mass 
functions $f(n)$ and $g(n)$
\cite{Bhattacharyya1943}, defined as
\[
B = \sum_i \sqrt{ f(i) \cdot g(i) }.
\]
This quantity is
a measure of the divergence between $f$ and $g$.

The flatness of an empirical histogram of $N$ discrete random variables
$X_1 \cdots X_N$ can be measured by replacing $f(n)$ by the sample
equivalent $\hat{f}(n) = \sum_i \frac{1}{N} h_n(X_i)$, where $h_n$ is
the indicator function of event $n$, and $g$ is the probability mass
function of the {\em null model}, i.e.\ $g(n) = \frac{1}{K}$ with $K$ being
the number of categories of the histogram. If $h_n$ would be fully
flat, one would obtain a BDM of $1$. 
In empirical data a BDM of $1$ is hardly reached and
deviations from $1$ of significant flat distributions depend on $K$ and $N$.

By randomization it is easy to assess
p-values, i.e.\ the probability that a BDM of $B_0$
or smaller occurred by pure chance under the assumption that the null
model is true \cite{Scott2004}: One generates
independent histograms with fixed $N$ and $K$ according to the null
model and counts the fraction of times, where the BDM is smaller than
the observed one. However if the p-values are very small one can
implement the method of Wilbur \cite{Wilbur1998} to save some
computation time.

Note that the BDM requires the empirical events to be
independent.  Hence, we generated histograms of independently visited ground
states for a small system ($L=40$) and low ground-state degeneracy
(we selected a realization exhibiting $K=16$) 
in the following way: For the optimized ensemble 
we considered the round-trip time $\tau$  and 
checked each $\tau$'th if  the random
walker sits currently at the ground state and, if so, the histogram
over all ground states is updated.
For simulated annealing, which provides a basis for ParQ,
this procedure  is not possible in this way, because there is no 
natural mixing time,
which could serve as a thinning interval.  Therefore we generated
histograms of {\em all} visited ground states and renormalized the
empirical histograms by considering an effective sample size such that
each annealing run has ``weight'' $1$, i.e.\ $h_n^{\text{eff}}(x) =
\frac{N_{\text{annealing}}}{N} h_n(x)$, where $N$ is the total number
of events.  

Note that in the case that the
random variable mixes faster than the number of walker's steps for one
annealing run the BDM for the effective histogram might be overestimated 
(and hence the p-values as well).
This would yield false positives.  However the opposite
case might not occur, because all annealing runs are independent.
Therefore the so defined effective histograms can only be used to
reject the model, which is exactly what we do here.

The results for both simulation methods are shown in
\Fig{fig:HIST}. The upper plot shows one of ten histograms of
independent runs for the optimized ensemble random walker, which has a
large p-value of $0.68$. The other nine runs yield p-values between
$0.007$ and $0.67$ ($0.4(1)$ on average) and hence we can accept the
null model.
For simulated annealing we find that not all ground states are visited
with equal probability (lower plot). The p-values for ten independent
runs varied between ($7\times 10^{-13}$ and $5\times
10^{-2}$). Therefore we have to reject the null model, hence
these simulated annealing runs visit ground states with a bias.
For an even faster cooling schedule we find p-values
between $2\times 10^{-24}$ and $2\times 10^{-3}$.  The reason for this 
bias might be
that the random walker gets stuck in preferable local minima. The
ground-state structure in form of a dendrogram illustrated in \fig{fig:HIST} 
below the histograms supports this argument.  The connection between two
ground states indicates that these two are merged into one cluster, and
the vertical distances are proportional to the Ward-distance, specified
in \Eq{eq:ward}.  One can see that \emph{within} the larges
 clusters the histogram becomes flatter and that the main source
of the non-flatness are differences in sampling \emph{between} the largest
clusters.

\section{Conclusion and Outlook} 

We investigated the relation between static and (MC) dynamic properties
of RNA secondary structures and the relation to the performance of
different MC algorithms. This model is an ideal test system
for this purpose for three reasons: i) The model exhibits
quenched disorder and
 has a complex low-energy landscape, where an interesting dynamical
behavior can be expected.
ii) It exhibits a static phase transition at finite temperature.
iii) The static behavior of the model can be analyzed exactly using
polynomial-time partition-function calculations for each single
realization of the disorder.

Analyzing the static behavior,  we calculated the DOS for
ensembles of sequences of different lengths.  In particular,
we studied the  ratio $g(E_1)/g(E_0)$, 
which plays the key role in the complexity of
MC methods. 
The distribution of this ratio could be fitted (but not perfectly) 
to a generalized extreme-value
distribution, similar as previously found for 
the case of $\pm J$ spin glasses.  Location, scale and
shape of this distribution scale algebraically with system size,
in contrast to $\pm J$ spin glasses.
We also computed Higgs's measure $s^\ast$ for the degree of 
ultrametricity of each realization
 and used hierarchical clustering approaches
to analyze the structure landscape.

For the dynamics, we examined three different MC approaches, which served as
basis for evaluating the infinite temperature transition matrix.
The nature of the model renders a direct MC implementation very
inefficient, hence we also included an $N$-fold way sampling
scheme.
Flat histogram and optimized ensemble methods provided examples
for equilibrium simulations, whereas ParQ is based on simulated
annealing, which is an out-of-equilibrium method.
In contrast to $\pm J$ model \cite{Heilmann2005} the ParQ method did not 
yield very accurate results near the ground
state and therefore equilibrium methods should be preferred.  
However, the disadvantage of theses methods is that
already suitable initial guesses of the DOS are required.  
Our simulations show that
ParQ can provide such guesses and hence a good strategy might be to
combine the ParQ method for a first estimate and use the optimized
ensemble method for further refinement.

The tunneling time, as a measure of complexity for the flat histogram
random walker, is also distributed according to a generalized 
extreme-value distribution. The scaling of location and scale parameters seems
to be algebraic with an exponent of $z \approx 7$, which differs from
the spin-glass model studied in the literature.  
The scaling of the shape parameter indicate
much larger sample to sample than the spin-glass case. Hence, computationally
very hard instances occur more often. 

Concerning the relation of static properties and
dynamical behavior, we found 
a strong correlation of the MC tunneling times to the value of the ratio 
$g(E_1)/g(E_0)$.
A similar correlation was found of the sampling error to $g(E_1)/g(E_0)$.
On the other hand, we could \emph{not} find a strong direct correlation
between MC tunneling times and degree of ultrametricity of the model.
Any numerically observed correlations appear only in a trivial way, i.e.\ due
to a correlation between the degeneracy ratio and $s^\ast$.
Hence, an ultrametric phase space (a kind of global characterization
of the energy landscape), as it seems to be present for
RNA secondary structures, does not necessarily lead to a complex
dynamics. The presence of meta-stable states, which is only a
local property of the energy landscape, appears to be much more important.

Finally we analyzed a possible reason of the failure of the ParQ
algorithm: Micro states with equal energy are not visited with equal
probability and hence evaluation of the infinite temperature
transition matrix does not work correctly.  This was not the case for
equilibrium methods.

Hence, we have seen that simple models of RNA secondary structures
are not only valuable for molecular-biology questions. They are also
in particular suitable
to study equilibrium and non-equilibrium properties of complex disordered systems and
their interdependence. 
Note that kinetic folding processes are by far more complex than the local 
update Monte Carlo algorithms used here \cite{Isambert2000} and a direct biological 
interpretation of our data is not possible.

Further studies in this field
will lead to a better understanding of fundamental 
equilibrium and non-equilibrium phenomena in disordered systems. 
Currently, we are also studying
aging phenomena for RNA secondary structures and we are hoping to understand them
better by comparing with the easily accessible equilibrium.

\section*{Acknowledgments}
The authors have received financial support from the {\em
  VolkswagenStiftung} (Germany) within the program ``Nachwuchsgruppen
an Universit\"aten'', and from the European Community via the
DYGLAGEMEM program. 
The simulations were performed
at the Gesellschaft f\"ur wissenschaftliche Datenverarbeitung mbH G\"ottingen 
and on a workstation cluster at the Institut f\"ur Theoretische
Physik, Universit\"at G\"ottingen, Germany.

\section*{References}
\bibliography{stefan}

\end{document}